\documentclass[prx, twocolumn,longbibliography]{revtex4-1}

\usepackage{bbm}
\usepackage{amsmath}
\usepackage{hyperref}
\usepackage{graphicx}
\usepackage{graphics}
\usepackage{amssymb}
\usepackage{mathtools}
\usepackage{xcolor}
\usepackage{physics}
\usepackage[british]{babel}
\usepackage{csquotes}
\usepackage{graphicx}
\usepackage{slashed}
\usepackage{float}

\graphicspath{{Figs/}}

\begin{document}


\title{Avoiding leakage and errors caused by unwanted transitions in Lambda systems}

\author{Arian Vezvaee$^{1,2}$} 
\author{Evangelia Takou$^1$} 
\author{Paul Hilaire$^1$} 
\author{Matthew F. Doty$^3$}
\author{Sophia E. Economou$^1$}
\affiliation{$^1$Department of Physics, Virginia Tech, Blacksburg, Virginia 24061, USA\\$^2$Department of Chemistry, University of Colorado at Boulder, Boulder, Colorado 80302, USA\\$^3$Department of Materials Science and Engineering, University of Delaware, Newark, Delaware 19716, USA}


\begin{abstract}
    Three-level Lambda systems appear in various quantum information processing platforms. In several control schemes, the excited level serves as an auxiliary state for implementing gate operations between the lower qubit states. However, extra excited levels give rise to unwanted transitions that cause leakage and other errors, degrading the gate fidelity. We focus on a coherent-population-trapping scheme for gates and design protocols that reduce the effects of the unwanted off-resonant couplings and improve the gate performance up to several orders of magnitude. For a particular setup of unwanted couplings, we find an exact solution, which leads to error-free gate operations via only a static detuning modification. In the general case, we improve gate operations by adding corrective modulations to the pulses, thereby generalizing the DRAG protocol to Lambda systems.  Our techniques enable fast and high-fidelity gates and apply to a wide range of optically driven platforms, such as quantum dots, color centers, and trapped ions.
\end{abstract}



\maketitle

\section{Introduction}

Quantum information processing requires the manipulation of quantum bits (qubits) via fast gates with high fidelities. Qubits are formed when a particular two-level subspace is chosen from a larger Hilbert space of a physical system. In specific cases, energy levels outside of the qubit subspace are used as an asset for auxiliary transfer of population within the qubit subspace and for the implementation of quantum gates. An important class of such setups are $\Lambda$-type systems that occur in several optically active qubit systems such as self-assembled quantum dots~\cite{Dutt,Economou2006,Economou2007,Donarini,Carter2021PRL,Vezvaee2021PRB}, nitrogen–vacancy (NV) centers~\cite{Kessler,Zhou2016nature,Zhou2017PRL,Yale2016nature}, trapped ions~\cite{Campbell,Toyoda,Takahashi,Duan_2001,Mizrahi2013PRL}, rare-earth ions~\cite{Kinos2021arxiv,Kinos2021PRA}, molecular qubits~\cite{Bayliss2020Science} and even in the microwave regime of superconducting circuits~\cite{Falci2013PRB,Earnest20108PRB}. Successful manipulation of $\Lambda$-systems is a key step to perform quantum information processing.

Various methods for the control of $\Lambda$-systems have been developed ~\cite{Economou2007,Economou2006,Shkolnikov,BaksicPRA,Ribeiro,shi2020,Gao,Moller,Eberly,Scala_2011}. However, in most platforms a bare three-level $\Lambda$-system is merely an idealization \cite{Wu}. Unintended interactions with other levels cause leakage and off-resonant couplings that are detrimental to the performance of quantum gates~\cite{Economou2012,Novikov}. The effect of these off-resonant unwanted couplings is intensified when the extra levels are close in energy to the auxiliary state. In principle, we can use longer pulses to resolve such small splittings. However, the qubit coherence time sets an upper bound to the duration of the pulse. Finding fast pulses that satisfy these opposing constraints and achieve high-fidelity gates is an open problem.

\begin{figure}
\includegraphics[scale=1]{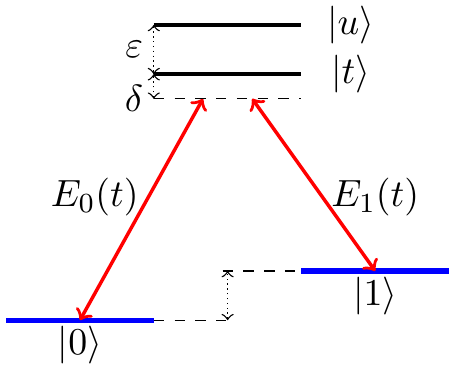}
\caption{(Color online)~Schematic depiction of a general $\Lambda$-type system with a fourth unwanted level. The qubit is defined in the subspace of the two lower levels $\{\ket{0},\ket{1}\}$. The coherent control is done through driving the transitions to the target level $\ket{t}$ (with the detuning $\delta$) which is separated from an unwanted level $ \ket{u}$ by the splitting $\varepsilon$. The off-resonant coupling to the unwanted level will cause low fidelities. We present resolutions for this problem by introducing a modification to the pulse and the detuning of the system.}
\label{fig-lambda-system}
\end{figure}

In this paper, we develop a novel framework relevant for systems with $\Lambda$-type selection rules, where the auxiliary excited state, hereafter referred to as the target state, is separated from an unwanted nearby level by a small energy splitting~(Fig.~\ref{fig-lambda-system}). The key insight of the presented formalism is that the unwanted level induces two different types of errors (leakage and phase errors), and that each is more prominent in a different regime. First, we consider the case that the target and unwanted states are formed from a set of two basis states. We show that in this case the unwanted couplings lead to phase errors, and the problem can be solved exactly. In this case, quantum gates with minimal gate error can be implemented through a modification of the detuning. Second, we consider the case that no basis state structure is present. We show that in this case the errors are in the form of leakage and we develop a novel version of the Derivative Removal by Adiabatic Gates (DRAG) framework for this system that battles the leakage error. Former versions of the DRAG method have been widely established as a powerful tool in dealing with unwanted dynamics and leakage cancellations in superconducting qubits~\cite{Motzoi2009,Motzoi2013,Gambetta2011,Theis2018,Werninghaus2021npj}. However, no version of this method applicable to $\Lambda$-type systems has yet been developed~\cite{Kinos2021arxiv}.

The paper is structured as follows: In Section~\ref{sec-system}, we present an overview of the $\Lambda$-system with unwanted transitions and discuss the relevance of the composition of the target and unwanted levels in terms of a set of two basis states. In Sec.~\ref{sec-exact-sol} we start with a hyperbolic secant uncorrected pulse and develop an exact analytical solution in the presence of such a basis state structure. In Sec.~\ref{sec-DRAG} we present the case without such a basis state structure, discuss the DRAG methodology, and develop a new DRAG-based formalism that resolves the off-resonant coupling issue in this case. In Sec.~\ref{sec-add-errors} We analyze the effects of additional errors, namely the cross-talk between the transitions of the $\Lambda$-system and spontaneous emission. We conclude in Sec.~\ref{sec-conclusion-drag}. The appendices contain the technical details of the CPT scheme and our DRAG formalism.


\begin{figure}
\includegraphics[scale=.35]{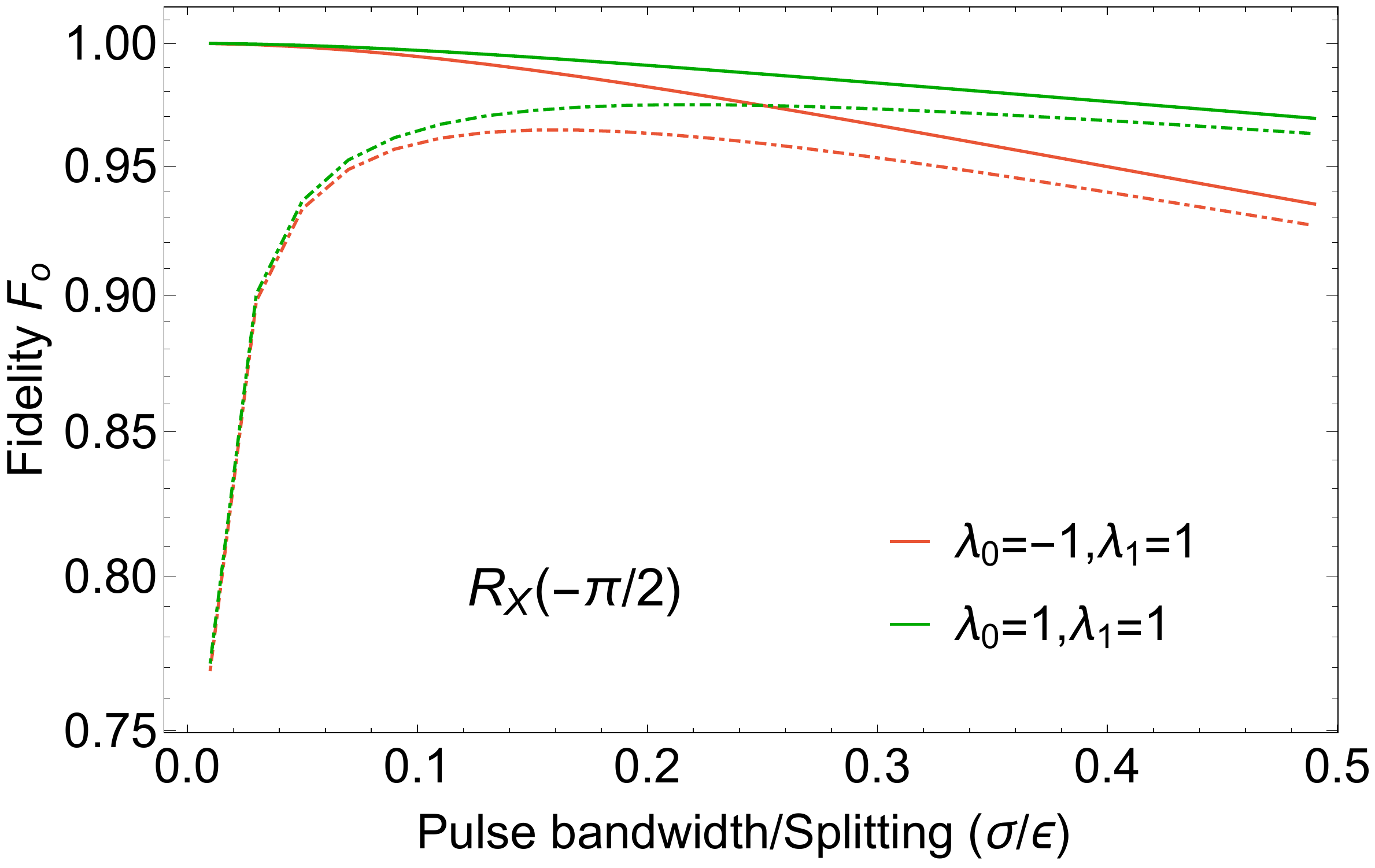}
\caption{(Color online)~Fidelity of an $X$-rotation by $-\pi/2$, $R_{X}(-\pi/2)$, in terms of the dimensionless parameter pulse bandwidth over splitting ($\sigma/\varepsilon$). Here, $\sigma$ is the bandwidth of the sech pulse and $\varepsilon$ is the splitting between the unwanted and target levels. The fidelities are shown for two cases of dependent ($\lambda_0=-\tan(\pi/4)=-1/\lambda_1$) and independent ($\lambda_0=1,~\lambda_1=1$) couplings. Without any corrective measures and ignoring spontaneous emission, reasonable fidelities required for quantum information processing are only achievable by using extremely narrow bandwidth pulses (solid lines). However, upon inclusion of spontaneous emission (dashed lines), even narrow bandwidths pulses will not lead to perfect fidelities.}
\label{fig-no-correction}
\end{figure}

\section{$\Lambda$-system with unwanted transitions under CPT} \label{sec-system}

In an ideal three-level $\Lambda$-system under CPT, the transitions are driven using two external fields (e.g. $E_0(t)$ and $E_1(t)$ in Fig.~\ref{fig-lambda-system}). Under the two-photon resonance (equal detuning of the two fields from the respective transitions they drive), destructive interference of the transitions caused by the two drive fields leads to the formation of a dark state, i.e. a state that is completely decoupled from the dynamics of the other two levels. This allows us to define the CPT frame, in which the system is described in terms of two new states in the qubit subspace, the dark state and its orthogonal bright state (denoted by $\ket{D}$ and $\ket{B}$, respectively), which are superpositions of the  $\ket{0}$ and $\ket{1}$ states.  
Then, the three-level system reduces to a two-level system where transitions are driven between the target excited state and the bright state. When we implement CPT with pulses that have hyperbolic secant (sech) envelopes, a choice that yields an analytically solvable time-dependent Schr\"odinger equation in a two-level system \cite{Rosen}, we can design arbitrary single-qubit rotations for the $\Lambda$-system~\cite{Economou2007}. The parameters of the sech pulses that drive the bright and target state can be chosen such that the evolution is \textit{transitionless}~\cite{Economou2006}: after the passage of a sech pulse, the population will always return to the bright state, with the bright state acquiring a non-trivial phase
\begin{equation}\label{eq-phase}
   \phi= 2 \arctan(\sigma/\delta),
\end{equation}
where $\sigma$ is the bandwidth, and $\delta$ is the detuning. Considering that the bright state spans the full Bloch sphere through the driving field parameters, this leads to full SU(2) qubit control. The technical aspects of the CPT framework and the sech-pulse control are provided in Appendix~\ref{app-cpt}. In this work, we are concerned with a non-ideal version of this scheme: a $\Lambda$-system with couplings to an additional, unwanted excited level.

The system under consideration is depicted in Fig.~\ref{fig-lambda-system}; the four states are comprised of a pair of low-energy levels $\{\ket 0 , \ket 1\}$ that encode the qubit, an auxiliary level $\ket{t}$ which is used to mediate the qubit rotation, and an unwanted level $ \ket{u} $. In contrast to the ideal CPT scheme, the additional excited level $\ket{u}$ (separated by an energy splitting $\varepsilon$ from the $\ket{t}$ state) introduces competing off-resonant couplings to the qubit states.  
Our goal is to perform the control of the qubit states using the target level in a way that eliminates or reduces the detrimental effect of the unwanted transition. Throughout this work, for off-resonant drivings, we take the frequency of the control pulses to be smaller than the transition frequency of the target transitions (i.e., negative detuning $\delta$); this choice minimizes the coupling to the unwanted level \footnote{Note that the presented analysis will be exactly the same if we swap the definition of the unwanted and target levels and change the sign of the detuning.}.

Figure ~\ref{fig-no-correction} shows the fidelity of a $-\pi/2$ rotation about the $x$ axis through direct application of the $\Lambda$-system CPT formalism where errors are caused by off-resonant coupling to the unwanted level. In the absence of corrective measures, high fidelity can be obtained only through the use of extremely narrow bandwidth pulses (thus using extremely long pulses in the time domain). In practice however, quantum gates should be implemented within the coherence time of the system and before it relaxes to its ground states through spontaneous emission. The relaxation time of relevant solid-state quantum emitters is usually short (e.g., $\sim$ one ns in quantum dots~\cite{Santori2004njp}, 1.85 ns in silicon vacancy~\cite{Becker2016NatComm},~4.5 ns in tin-vacancy~\cite{Trusheim2021PRL}, and 10 ns in NV centers~\cite{Jelezko2006}), thus requiring broad bandwidth pulses in general. We develop a formalism that deals with the issue of off-resonant couplings without trading off the duration of the gates for selectivity, ensuring operations performed within the relaxation time. The effects of spontaneous emission are further discussed in Section~\ref{sec-se}.

As mentioned above, the two legs of the $\Lambda$-system are distinct and each transition is driven by a single drive field ($E_0(t)$ or $E_1(t)$), as shown in Fig.~\ref{fig-lambda-system}. This can be satisfied by either polarization selection rules or large energy separation of the ground states, depending on the specifics of the platform considered. In the former case, the orthogonality of each transition dipole with one drive field ensures that each transition couples to a single drive. In the latter case, sufficient energy separation of the ground-state levels implies that the off-resonant couplings of the drive fields to the opposite $\Lambda$-transitions average out. We relax this assumption and discuss the effects of the cross-talk in Section~\ref{sec-crosstalk}.

We consider two separate cases for the composition of the target and unwanted levels that lead to different physics of the system. First, the case that the two excited states, $|t\rangle$ and $|u\rangle$, are formed from superpositions of two basis states,  $|b_0\rangle$ and $|b_1\rangle$: $|t\rangle=\sin(\eta)|b_1\rangle - \cos(\eta)|b_0\rangle$ and $|u\rangle = \cos(\eta) |b_1\rangle +\sin(\eta)|b_0\rangle$, where $\ket{0}$ couples only to $\ket{b_0}$, and $\ket{1}$ couples only to $\ket{b_1}$, with dipole moments $d_{0,b_0}$ and $d_{1,b_1}$, respectively. In this case, the parameter $\eta$ determines the weight of the composition of the two basis states. Since $\ket{0}$ couples only to $\ket{b_0}$, we have the Rabi frequency $\Omega_0(t) \equiv -\cos(\eta) d_{0,b_0} E_0$. Similarly, since $\ket{1}$ couples only to $\ket{b_1}$ the Rabi frequency $\Omega_1(t) \equiv \sin(\eta) d_{1,b_1} E_1$. This choice translates into a set of couplings to the unwanted level given by $\lambda_0$ (for the $|0\rangle \leftrightarrow |u\rangle$ transition driven by $E_0$) and $\lambda_1$ (for the $|1\rangle \leftrightarrow |u\rangle$ transition driven by $E_1$), that are inversely related: $\lambda_0\equiv -\tan(\eta) = -1/\lambda_1$.

Alternatively, we could consider the case that the target and unwanted level have bare couplings to the two qubit states. In this case we can set the couplings to the target state to unity (i.e., $\Omega_0(t) \equiv d_{0,t} E_0$ and  $\Omega_1(t) \equiv d_{1,t} E_1$) and take the couplings to the unwanted state to be proportional to $\lambda_0$ and $\lambda_1$ as the \textit{bare} coupling: $\lambda_0 \Omega_0(t)$ (for the $|0\rangle \leftrightarrow |u\rangle$ transition) and $\lambda_1 \Omega_1(t)$ (for the $|1\rangle \leftrightarrow |u\rangle$ transition). As such, the $\lambda_0$ and $\lambda_1$ parameters (coupling strengths) will be independent of each other. 

In the following we discuss the errors caused by each of the cases above and develop methods that allow implementation of fast and high-fidelity gates. We leave the numerical simulations regarding the nature of the errors in each case (that is, phase error versus leakage), until Section~\ref{sec-conclusion-drag} where we have developed enough methodology. In Section~\ref{sec-exact-sol}, we discuss that the errors from the case of dependent couplings ($\lambda_0\equiv -\tan(\eta) = -1/\lambda_1$) are mostly phase errors and we can achieve quantum gates with negligible error simply by modifying the drive frequency. We discuss the case of independent couplings in which the errors are due to leakage in Section~\ref{sec-DRAG} where we develop a novel version of the DRAG method to achieve high fidelity gates in these systems.

\begin{figure}
\includegraphics[scale=1.2]{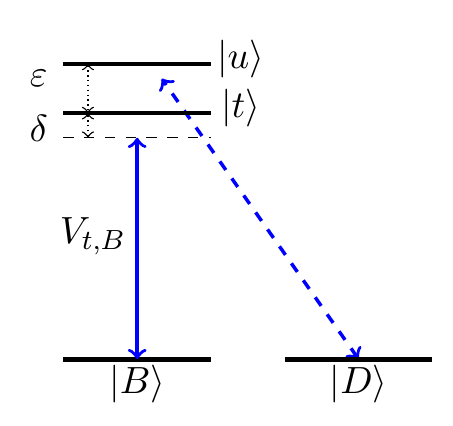}
\caption{(Color online)~Selection rules in the dressed basis of CPT in the case of equal couplings, i.e., $\eta=\pi/4$. In this case the problem reduces to two dissociated two-levels (bright and target, and dark and unwanted), each subject to a transitionless pulse. }
\label{fig-cpt}
\end{figure}

\section{Dependent couplings: An exact solution approach }\label{sec-exact-sol}

In this section, we first consider the $\Lambda$-system with an additional excited state formed from basis states that lead to dependent couplings. The additional excited state induces off-resonant couplings outside of the $\Lambda$-subspace, which in the CPT frame translates into transitions that link the bright and dark states to the unwanted level. In an ideal $\Lambda$-system, one would drive the target transitions with the fields $E_\ell(t)=\Omega_{\ell,o}(t)\cos(\omega_{\ell} t)$ ($\ell=0,1$), and static detuning $\delta$ to implement the desired gate operation. Here $o$ refers to the original drive fields, described by the unperturbed hyperbolic secant envelopes, and $\omega_l$ is the drive frequency. The resonant frequencies for the transitions $|0\rangle \leftrightarrow |t\rangle$ and $|1\rangle \leftrightarrow |t\rangle$ are denoted as $\omega_{0t}$ and $\omega_{1t}$ respectively; the detuning $\delta$ is defined as $\delta=\omega_0-\omega_{0t}=\omega_1-\omega_{1t}$. In the following, as is standard for CPT, we make use of the rotating wave approximation (RWA), i.e., we define $E_\ell(t)= \text{exp}(-i\omega_l t) \Omega_{\ell,o}(t)+\text{c.c.}$~.

Although our formalism is general and applicable to the design of arbitrary axis single-qubit gates, we choose to showcase our protocols in the particular context of $X$-gates (or equivalently $Y$-gates, up to a phase between the two drives). For this reason, we set the two Rabi frequencies of the target transitions (i.e. $|0\rangle \leftrightarrow |t\rangle$ and $|1\rangle \leftrightarrow |t\rangle$) to be equal, that is $\Omega_o\equiv{\Omega}_{0,o}={\Omega}_{1,o}$. Under this condition and RWA, the Hamiltonian in the CPT frame rotating with the drive frequency is given by (see Appendix~\ref{app-cpt} for derivation):
\begin{eqnarray} \label{eq-X-H}
H_{\text{CPT}}&=&(\delta/2) (\Pi_D +\Pi_B-\Pi_t-\Pi_u)+\varepsilon \Pi_u \nonumber  \\
&&+~ \big\{ \sqrt{2}\Omega_o |B\rangle\langle t| 
+ \frac{1}{\sqrt{2}}\Omega_o (\lambda_0 -\lambda_1) |D\rangle\langle u| \nonumber \\ &&+~ \frac{1}{\sqrt{2}} \Omega_o (\lambda_0 +\lambda_1) |B\rangle\langle u|+ \text{h.c.}\big\},
\end{eqnarray} where $\Pi_m = \op{m}$. Notice that the CPT Hamiltonian above is generic since the relation among the $\lambda_0$ and $\lambda_1$ have been kept implicit. However, in this section we consider the case of dependent couplings: $\lambda_0\equiv -\tan(\eta) = -1/\lambda_1$. In this case, when $\eta=\pi/4$ (equivalently $\lambda_0=-1,~\lambda_1=1$), the four-level system in the CPT frame transforms into two independent two-level systems, each driven by a sech pulse (Fig.~\ref{fig-cpt}): the bright and target, and the dark and unwanted. For a single two-level system, the population returns to the ground state at the end of the evolution if the absolute value of the Rabi frequency is equal to the bandwidth of the pulse (see Appendix~\ref{app-cpt}). For $\eta=\pi/4$ this condition is satisfied for both two-level systems. Under this condition, we can therefore solve analytically the problem as it separates into two separate two-level problems. This enables us to find an exact solution, where the unwanted level is fully taken into account and its dynamics are incorporated into the gate design.

\begin{figure}[!htbp]
\includegraphics[scale=.37]{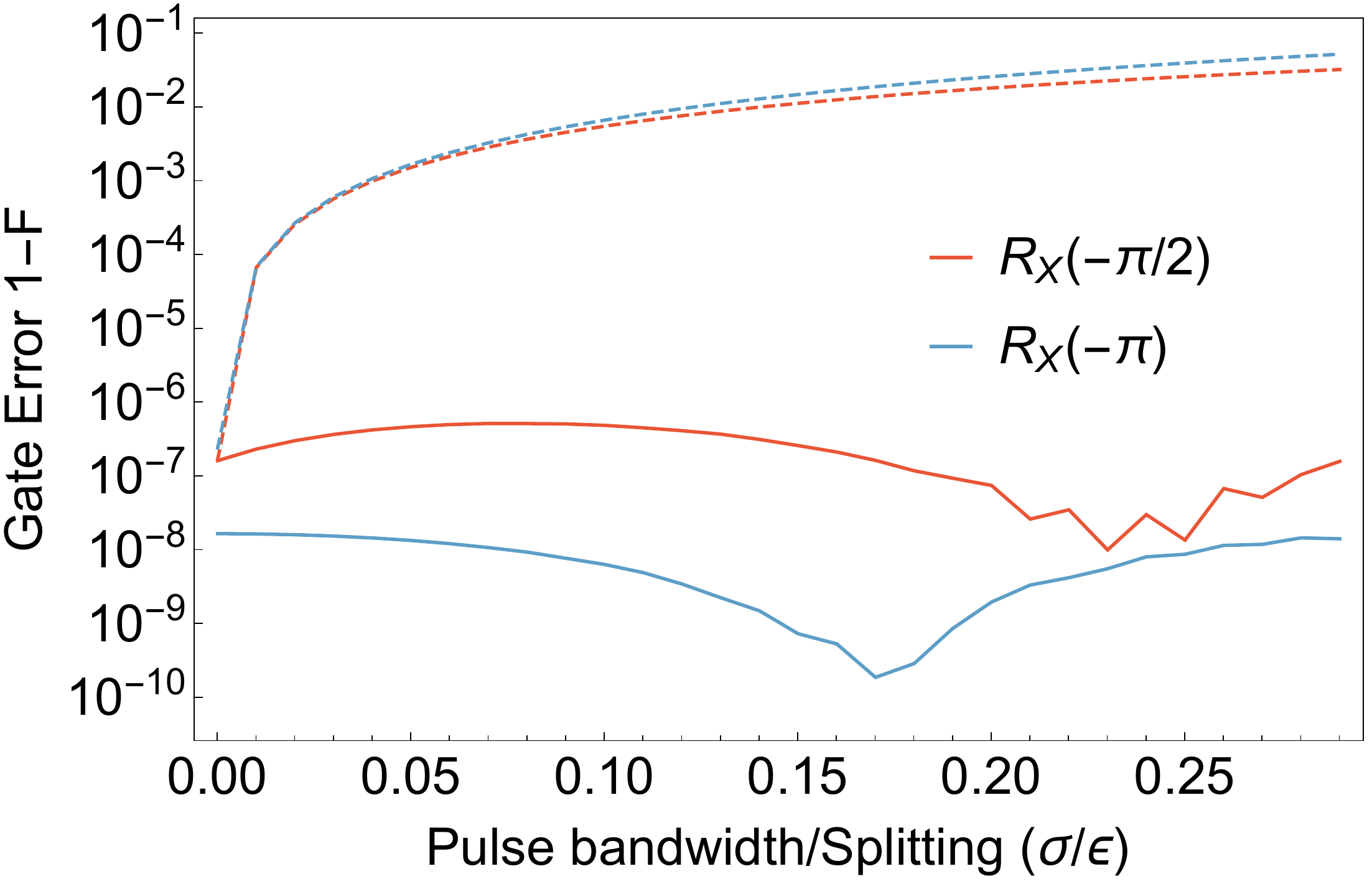}
\caption{(Color online)~$R_{X}(-\pi/2)$ and $R_{X}(-\pi)$ rotations for pulses with no correction (dashed) and exact modifications of detuning (solid) given by Eq.\eqref{eq-exact}, in terms of the dimensionless parameter bandwidth over splitting ($\sigma/\varepsilon$). The gate times of the sech pulses are chosen to be $16/\sigma$ and the non-monotonic behavior of $R_{X}(-\pi)$ is due to negligible numerical instability rising from cutting off the sech pulses. The exact approach enables gate implementation with negligible gate error. }
\label{fig-exact}
\end{figure}

The difference in the dynamics of the two two-level systems arises from the different detunings; each ground state acquires a different phase at the end of the evolution. Thus, the total phase (rotation angle in the bright-dark subspace) will be $\phi_\text{tot} = \phi_{B,t} - \phi_{D,u}$, where $\phi_{B,t} = 2\arctan(\sigma/\delta)$, and $\phi_{D,u} = 2\arctan(\sigma/(\delta-\varepsilon))$. Therefore the error in implementation of the desired gate is caused by the deviation of $\phi_\text{tot}$ from the intended phase in the bright-dark subspace. We can account for this phase difference by writing $\phi_\text{tot}$ as  
\begin{equation}
    \phi_\text{tot} = 2\arctan(\frac{-\sigma \varepsilon}{(\delta-\varepsilon)\delta + \sigma^2}),
\end{equation}
and then obtain the detuning modification required for the exact implementation of any desired rotation angle. To implement a rotation by $\phi_\text{tot}=-|\phi|$, we find
\begin{equation} \label{eq-exact}
    \delta   =  \frac{1}{2}(\varepsilon \pm \sqrt{\varepsilon^2 + 4 \varepsilon\sigma\cot\big[\frac{|\phi|}{2}\big]} -4 \sigma^2.
\end{equation}
By setting the detuning to either branch of the equation above, a $-|\phi|$ rotation with negligible gate error will be implemented. For instance, $R_{X}(-\pi/2)$ and $R_{X}(-\pi)$ are illustrated in Fig~\ref{fig-exact}. These gate performances are quantified by averaging over all input states existing in the Hilbert space, which can be reduced to the following expression~\cite{Bowdrey}:
\begin{equation}
	\begin{split}\label{eq-fidelity}	
		F_i	=&\frac{1}{6}\sum_{j=\pm x, \pm y, \pm z}\mathrm{Tr}\left[U_\mathrm{ideal} \rho_j U_\mathrm{ideal}^\dagger \mathcal{U}_i(\rho_j)\right].
	\end{split}
\end{equation}
Here, the $\rho_j$'s are the six cardinal states on the Bloch sphere, $\mathcal{U}_i(\rho_j)$ is the evolution of the axial vectors under the actual evolution of the system, and $i$ is either the original or the exact solution. The ideal evolution in the subspace of $\{\ket D, \ket B\}$ is given by $    U_{\text{ideal}}=\text{diag}(e^{- i\phi },e^{ i\phi })$, where $\phi$ is given by Eq.~\eqref{eq-phase}. We will use this quantification of the gate performances throughout the rest of this paper.

\begin{figure*}[!htbp]
\includegraphics[scale=.7]{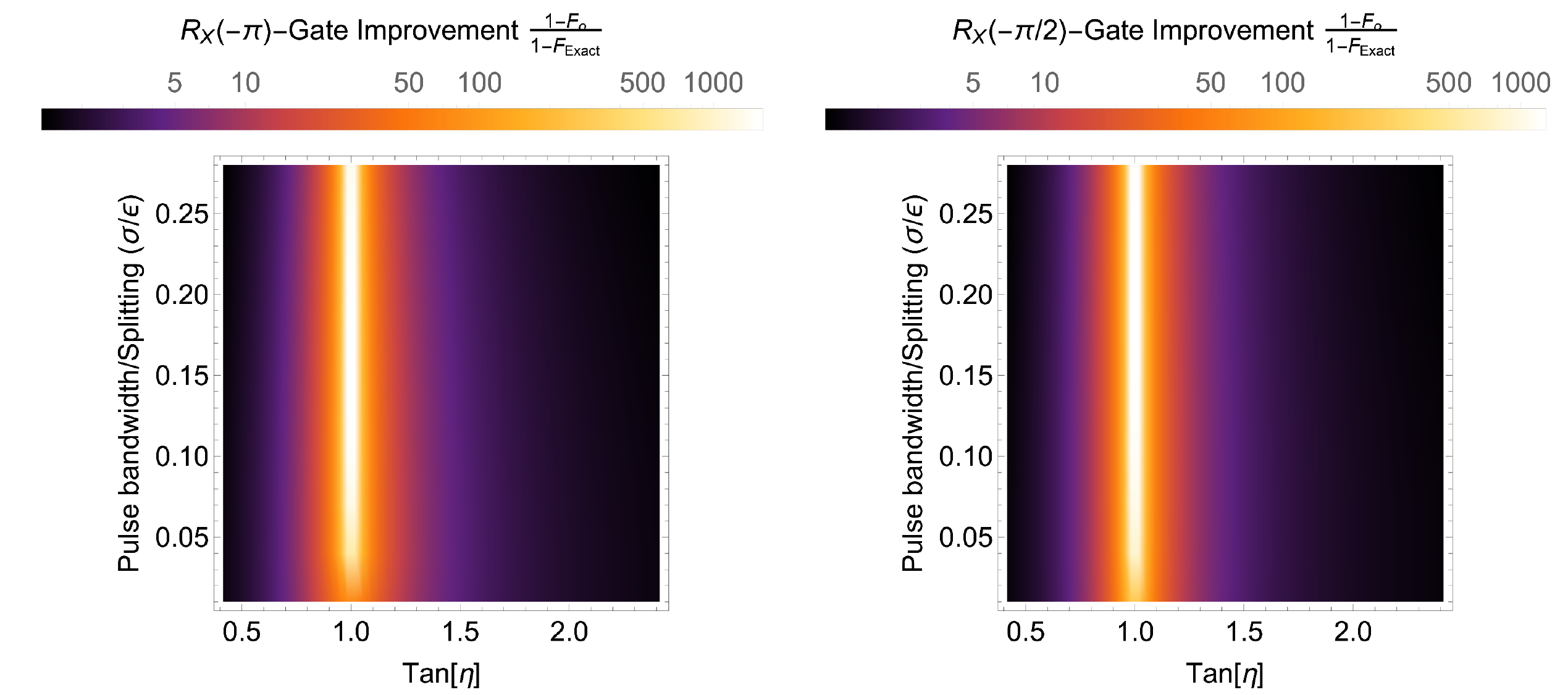}
\caption{(Color online)~Gate improvement of (a) $R_{X}(-\pi)$ and (b) $R_{X}(-\pi/2)$, in terms of the dimensionless parameter pulse bandwidth over splitting ($\sigma/\varepsilon$), and the parameter $\eta$ which determines the weight distributions of the basis states in the target and unwanted levels $\ket t = \sin(\eta) \ket{b_1} -\cos(\eta) \ket{b_0}$ and $\ket u = \cos(\eta) \ket{b_1}+\sin(\eta)\ket{b_0}$. The best improvements occur for when the two dissociated two-level approximation is valid ($\eta\approx\pi/4$). However, even away from this value, the exact solution leads to substantial gate improvements, through a single modification of the detuning.}
\label{fig-exact-improvment}
\end{figure*}

It is noteworthy that, as Fig.~\ref{fig-exact} shows, without pulse shaping/chirping, or increased experimental overhead, we can realize unit fidelity operations by incorporating the unwanted level into our quantum control design. As we move further away from this particular setup of coupling strengths, the two two-level systems become coupled. Nevertheless, our analytic solution still gives rise to substantial gate improvement. 
We define the gate improvement as the ratio of the original gate error to the gate error using the improved solution. This is shown in Fig.~\ref{fig-exact-improvment}. As seen from the figure, away from $\eta=\pi/4$ the gate performance improves, with of course the best improvement occurring near $\eta\approx\pi/4$ where the two two-level system formation of CPT frame is valid.

The two two-level systems formation  does not necessarily occur in the case of independent couplings. More generally, if the two couplings have the same sign, then the errors will not be in the form of phase errors anymore. For this reason, when the couplings are independent we resort to a different strategy based on perturbative expansion methods, as we describe in the following section.

\section{Independent couplings: DRAG formalism with CPT }\label{sec-DRAG}

In this section, we consider the case of a $\Lambda$-system with an additional unwanted level, where the couplings to the auxiliary states are independent of each other but have the same sign. To gain insight into how the CPT-transformed Hamiltonian (Eq.~\eqref{eq-X-H}) in the system would look in a limiting case, we may consider when the couplings are equal to each other and have the same strength as the coupling to the target level: $\lambda_0=\lambda_1=1$. It can be seen from Eq.~\eqref{eq-X-H} that in this case the system turns into a V-type system rather than two dissociated two-level systems. The errors, in this case, will not be a simple phase error that can be taken care of by modification of the detuning. In fact, as we will demonstrate in Section~\ref{sec-conclusion-drag}, the source of errors, in this case, is leakage to the unwanted level. Therefore, we develop a novel version of the DRAG method tailored to CPT that allows the implementation of high-fidelity gate operations.

To overcome the errors for the present case via the DRAG approach, we start by modulating the original pulses. We consider an additional corrective drive $\Omega_{l,c}(t)$ ($\ell=0,1$) for each of the two fields, phase detuned from the original drive by $\pi/2$. We further set the frequency to be the same as that of the original drive, hence reducing the experimental overhead of an additional laser drive. From here on, we use the letter $c$ to refer to any subsequent parameters of the corrective drive fields. We choose the total fields of the system to be $E_\ell(t)=(1/2) \Omega_{\ell,o}(t) \cos(\omega_{\ell} t)+(1/2) \Omega_{\ell,c}(t) \sin(\omega_{\ell} t)$ which in RWA are equivalent to $E_\ell(t)=(1/2) \text{exp}(-i\omega_l t) (\Omega_{\ell,o}(t)+i\Omega_{\ell,c}(t))+\text{c.c.}$
Following the previous section, we apply our formalism to the context of $X$-rotations, so we set the two Rabi frequencies of the target transitions to be equal, that is $\Omega_o\equiv{\Omega}_{0,o}={\Omega}_{1,o}$, with the additional condition of $\Omega_c\equiv{\Omega}_{0,c}={\Omega}_{1,c}$. However, we note that these conditions can be lifted and the formalism we develop remains valid, with the difference that the rotation axis changes.

Under these conditions and after performing the RWA, the Hamiltonian in the CPT frame is given by (see Appendix~\ref{app-cpt})
\begin{equation}\label{eq-X-H-DRAG}
\begin{split}
    \tilde{H}_{\text{CPT}}&=(\delta/2)(\Pi_B+\Pi_D-\Pi_t-\Pi_u)+\varepsilon\Pi_u \\&~+ \frac{1}{2\sqrt{2}}\Big\{ 2\sum_{j=o,c}\Omega_j\sigma^j_{B,t}  \\&~+(\lambda_0-\lambda_1)\sum_{j=o,c}\Omega_j\sigma^j_{D,u} \\&~+(\lambda_0+\lambda_1)\sum_{j=o,c}\Omega_j\sigma^j_{B,u} \Big \},
    \end{split}
\end{equation}
where to indicate the matrix elements for the generic transition $\ket m\xleftrightarrow[]{}\ket n $, we have defined $\sigma_{m,n}^o \equiv \ket{n}\bra{m}+\ket{m}\bra{n}$ and $\sigma_{m,n}^c \equiv i \ket{n}\bra{m}-i\ket{m}\bra{n}$. 
Previous formulations of the DRAG method have focused on canceling out leakage errors in ladder-type systems (e.g., transmons) that occur between consecutive energy levels, making it inapplicable to $\Lambda$-systems. The complexity, in this case, arises as the qubit control is performed indirectly via the target level. Therefore, to mitigate the unwanted transitions of $\Lambda$-type structures we modify the DRAG method as we explain below.

Under the DRAG formalism, corrections to the driving fields are determined by utilizing a frame transformation~\cite{Theis2018}. In this new DRAG frame, the resulting Hamiltonian is constrained such that it implements the target evolution. The DRAG frame Hamiltonian for a generic Hamiltonian $H(t)$, is generated by the transformation $A(t)=e^{-iS(t)}$ and is given by
\begin{equation}\label{Eq7}
    H_{\text{DRAG}}=A^\dagger(t)H(t)A(t) + i \dot{A}^\dagger(t) A(t).
\end{equation}
In our case, $H(t)\equiv  \tilde{H}_{\text{CPT}}(t)$.
The operator $S(t)$ can be an arbitrary Hermitian operator as long as it respects the boundary conditions of the transformation. That is, the frame transformation has to vanish at the beginning and end of the pulse ($A(0)=A(t_g)=\textbf{1}$), such that the ideal gate we wish to design remains the same in both the CPT and DRAG frames. The generality of $S(t)$ would in principle generate a wide range of pulse corrections. However, extracting such closed-form expressions is non-trivial for our four-level system (as the pulses are time-dependent), hence, we turn to a perturbative expansion of the transformation. To that end, we utilize the Schrieffer-Wolff (SW) transformation~\cite{Schrieffer} and its perturbative expansion. 

Additionally, we expand the CPT Hamiltonian of Eq.~\eqref{eq-X-H-DRAG} into a power series. To do so, we first make the CPT Hamiltonian dimensionless by multiplying all quantities by the gate time $t_g$ \cite{Gambetta2011}. The Hamiltonian is then expanded into a power series (Appendix~\ref{app-derivations}) of $x=1/\varepsilon t_g$, where $\varepsilon$ is the energy cost by which the unwanted subspace of $|u\rangle$ is off-resonant with respect to the $\Lambda$-system. 
This way, one can collect the same orders of the expansion in the left-hand side and right-hand side of Eq.~\eqref{Eq7} and relate the elements of $H_{\text{DRAG}}$ to the corrective fields (involved in $\tilde{H}_{\text{CPT}}$) and to elements of $S(t)$ order by order. Further details of this procedure are given in Appendix~\ref{app-derivations}.

\begin{figure}
\includegraphics[scale=.4]{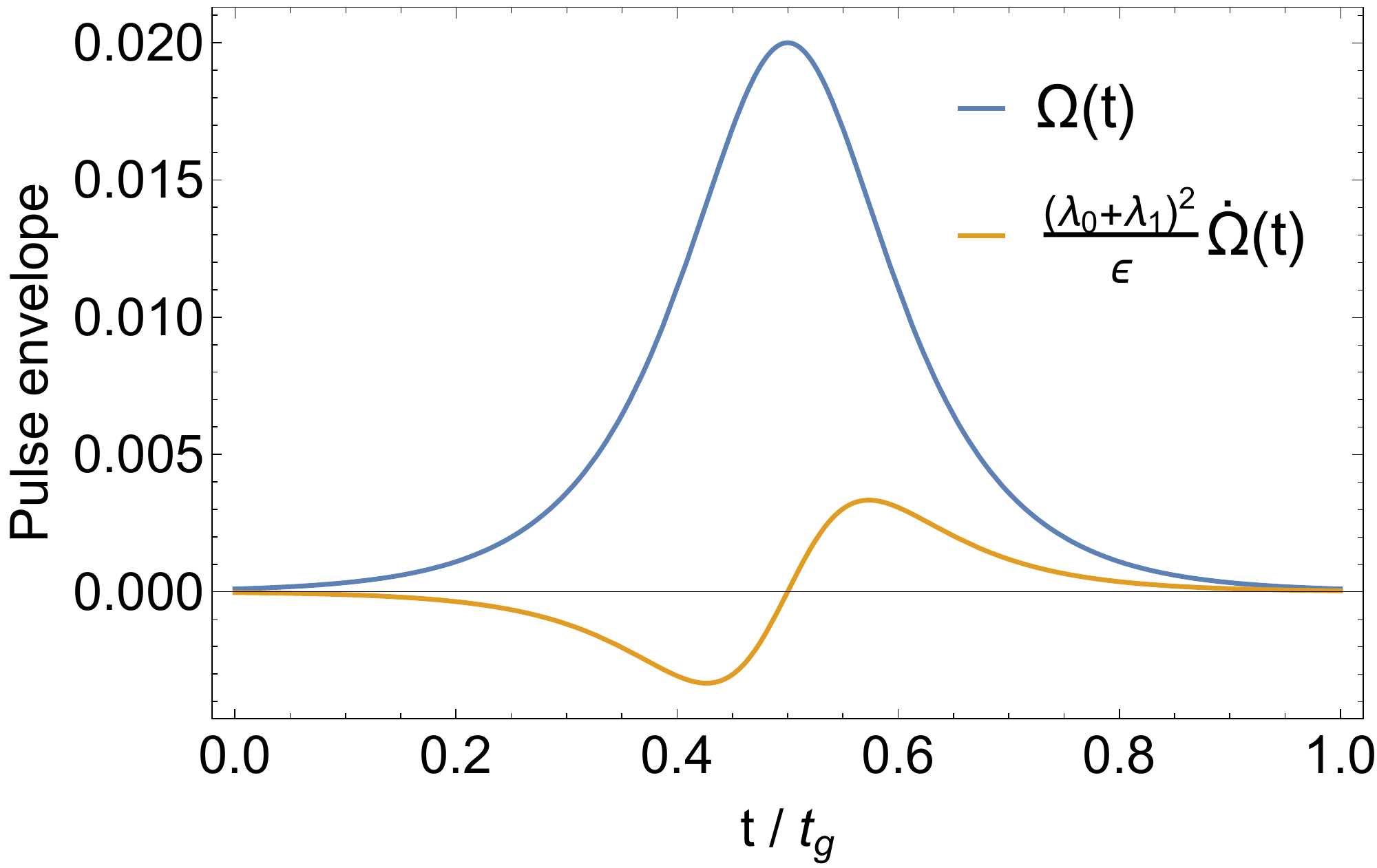}
  \caption{(Color online)~An example of  pulse shapes for a system with $\eta=\pi/3$ and splitting of $\varepsilon=80$~$\mu$eV. The sech pulse (in blue) is $\Omega(t) = \sigma \sech(\sigma (t-t_g/2))$, and its derivative corrective solution (in orange) modulated with the splitting and the couplings is $\Omega(t) = (\lambda_0+\lambda_1)^2(1/\varepsilon) \frac{d}{dt}\sigma \sech(\sigma (t-t_g/2))$. The bandwidth is taken to be $\sigma=0.02$~meV, and gate time is $t_g=16/\sigma$.  }
\label{fig-pulse-shape}
\end{figure}

Our goal is to constrain the DRAG-frame Hamiltonian such that it implements our target evolution. To this end, we define a target Hamiltonian, capable of performing arbitrary rotations within the qubit (dark-bright) subspace,
\begin{equation}
 H_\mathrm{target}^{\mathrm{CPT}}(t)=  \frac{1}{2}\sum_{i=o,c} h_{i}(t)\sigma_{B,t}^i  +\frac{1}{2}h_z(t) \left(\Pi_B - \Pi_t\right),
\end{equation}where $h_{i}(t)$ and $h_z(t)$ are arbitrary control fields. Here we choose the $h_{i}(t)$ to be sech pulses, where $h_z(t)$ corresponds to the detuning. To ensure that the DRAG frame Hamiltonian implements the intended operation as dictated by $H^{\text{CPT}}_{\text{target}}$, we impose the following constraints:\begin{eqnarray} \label{eq-X-control}
&&\hspace{-5mm}\mathrm{Tr}[ H_\mathrm{DRAG}(t) \sigma^i_{B,t}]= h_i(t),  \\
&&\hspace{-5mm}\mathrm{Tr}[ H_\mathrm{DRAG}(t)  \left(\Pi_B - \Pi_t \right)]= h_z(t), \nonumber\end{eqnarray}where $i\in\{o,c\}$. Notice that these constraints ensure that the zeroth order DRAG Hamiltonian remains the same as the ideal Hamiltonian of the CPT frame. Additionally, to enforce decoupling from the unwanted subspace in the new Hamiltonian, we impose the following constraints:
\begin{eqnarray} \label{eq-X-decoupling}
  \mathrm{Tr}[ H_\mathrm{DRAG}(t) \sigma^i_{D,u}]&=&0, \enskip  \nonumber \\
  \mathrm{Tr}[ H_\mathrm{DRAG}(t) \sigma^i_{B,u}]&=&0, \\ \enskip \mathrm{Tr}[ H_\mathrm{DRAG}(t) \sigma^i_{t,u}]&=&0, \nonumber \\\nonumber
\end{eqnarray}where $i\in\{o,c\}$. The constraints can be solved consistently and lead to a set of corrective pulses for each order of the expansion. The details of the derivation for the corrective fields can be found in Appendix~\ref{app-derivations}. To the first order of expansion, for the two drives being subjected to an uncorrected pulse $\Omega(t)$, the simplest solution which respects the boundary conditions of the transformation is given by
\begin{eqnarray} 
\Omega_{0,o} (t)&=&\Omega_{1,o} (t)= \sqrt{2}\left(1+\delta\frac{(\lambda_0+\lambda_1)^2}{16\varepsilon}\right) \Omega(t), \label{eq-DRAG-sols-o} \\
\Omega_{0,c} (t)&=&\Omega_{1,c} (t)= \frac{\sqrt{2}}{16\varepsilon}(\lambda_0+\lambda_1)^2 \dot\Omega(t). \label{eq-DRAG-sols-c}
\end{eqnarray}
We depict the sech pulse envelopes for one specific set of parameters in Fig.~\ref{fig-pulse-shape}.

Due to the indirect control of the qubit in $\Lambda$-systems, the diagonal constraint of Eq.~\eqref{eq-X-control}, unlike what happens in transmon qubits, does not lead to a global phase among all states for the first-order DRAG Hamiltonian. Instead, it resolves the off-resonant coupling at the cost of inducing a phase between the bright and dark states. Therefore, we need to investigate the form of the first-order DRAG Hamiltonian to infer this phase between the dark and bright states. By restricting our attention to the $\Lambda$-system subspace, we find the zeroth-, and first-order DRAG frame Hamiltonian in the subspace of $\{\ket{D},\ket{B},\ket{t}\}$ to be

\begin{equation} \label{eq-zero-order-drag}
   H_\mathrm{DRAG}^{(0)}=\begin{bmatrix}
\delta/2 & 0 & 0 \\
0 & \delta/2 & \Omega(t)  \\
0 &\Omega(t) & -\delta/2
\end{bmatrix},
\end{equation}
and 
\begin{widetext}
\begin{equation} \label{eq-first-order-drag}
     H_\mathrm{DRAG}^{(1)}=\begin{bmatrix}
-\frac{1}{8\varepsilon}\left(\lambda_{0}-\lambda_{1}\right)^{2}\Omega^{2}(t) & \frac{1}{8\varepsilon}\left(-\lambda_{0}^{2}+\lambda_{1}^{2}\right)\Omega^{2}(t) & 0 \\
\frac{1}{8\varepsilon}\left(-\lambda_{0}^{2}+\lambda_{1}^{2}\right)\Omega^{2}(t) & -\frac{1}{16\varepsilon}\left(\lambda_{0}+\lambda_{1}\right)^{2}\Omega^{2}(t) & 0   \\
0 & 0 & -\frac{1}{16\varepsilon}\left(\lambda_{0}+\lambda_{1}\right)^{2}\Omega^{2} (t)
\end{bmatrix}.
\end{equation}
\end{widetext}

Notice that for the limiting case of a V-system (i.e., $\lambda_0=\lambda_1=1$), all elements of $H_{\text{DRAG}}^{(1)}$ in Eq.~\eqref{eq-first-order-drag} vanish, except for the diagonal entry of the bright state. While the zeroth-order DRAG Hamiltonian stays in the form enforced by Eqs.~\eqref{eq-X-control}, we note that in the first-order DRAG Hamiltonian there is an induced phase between the dark state and the bright-target subspace. In this Hamiltonian (Eq.~\eqref{eq-first-order-drag}), the bright and target states follow the same phase evolution, as dictated by the common diagonal element $-1/(16\varepsilon)(\lambda_0+\lambda_1)^2\Omega^2(t)$. However, the qubit states are composed of the dark and bright states, which up to the first order (neglecting the off-diagonal elements of $H_{\text{DRAG}}^{(1)}$) evolve with a different phase. Effectively, this implies that the DRAG corrections reduce the unintended off-resonant couplings to the unwanted level at the cost of inducing a relative phase between the qubit states in the CPT frame. This is an immediate consequence of the fact that we counteract the unwanted couplings indirectly; in the CPT frame, we design a target Hamiltonian that involves transitions between the bright and target levels.

\begin{figure*}
\includegraphics[scale=.7]{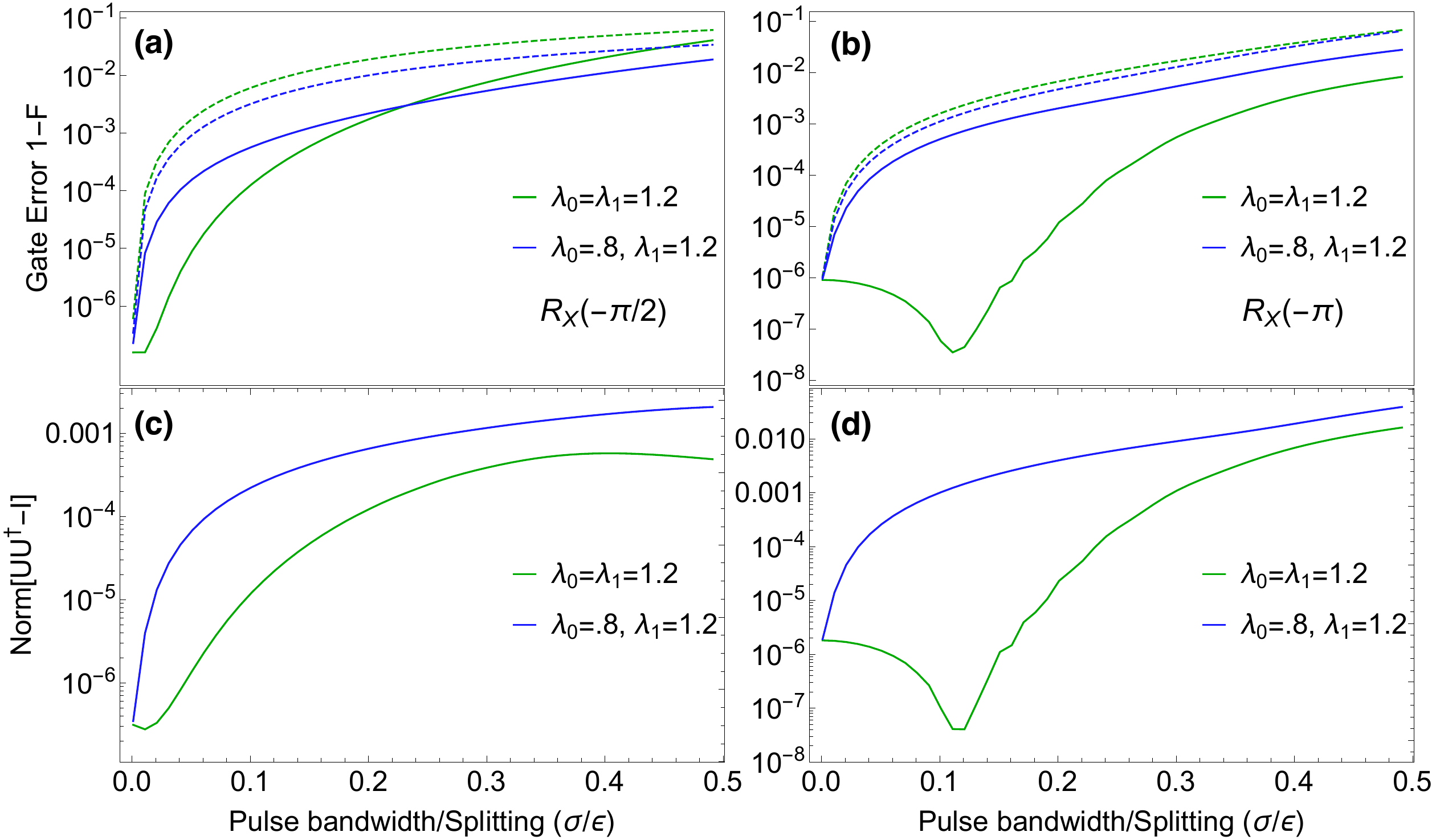}
\caption{(Color online)~Gate error comparison in terms of the dimensionless parameter pulse bandwidth over splitting ($\sigma/\varepsilon$) for the original pulse (dashed) and DRAG solution (solid) for (a) $-\pi/2$ and (b) $-\pi$ rotation about the $x$ axis. (c) and (d) are the corresponding deviation from unitarity ($|UU^\dagger-\textbf{1}|$) of (a) and (b), respectively. The colors correspond to different strengths of couplings to the unwanted level.}
\label{fig-comp-x}
\end{figure*}

\begin{figure*}
\includegraphics[scale=.67]{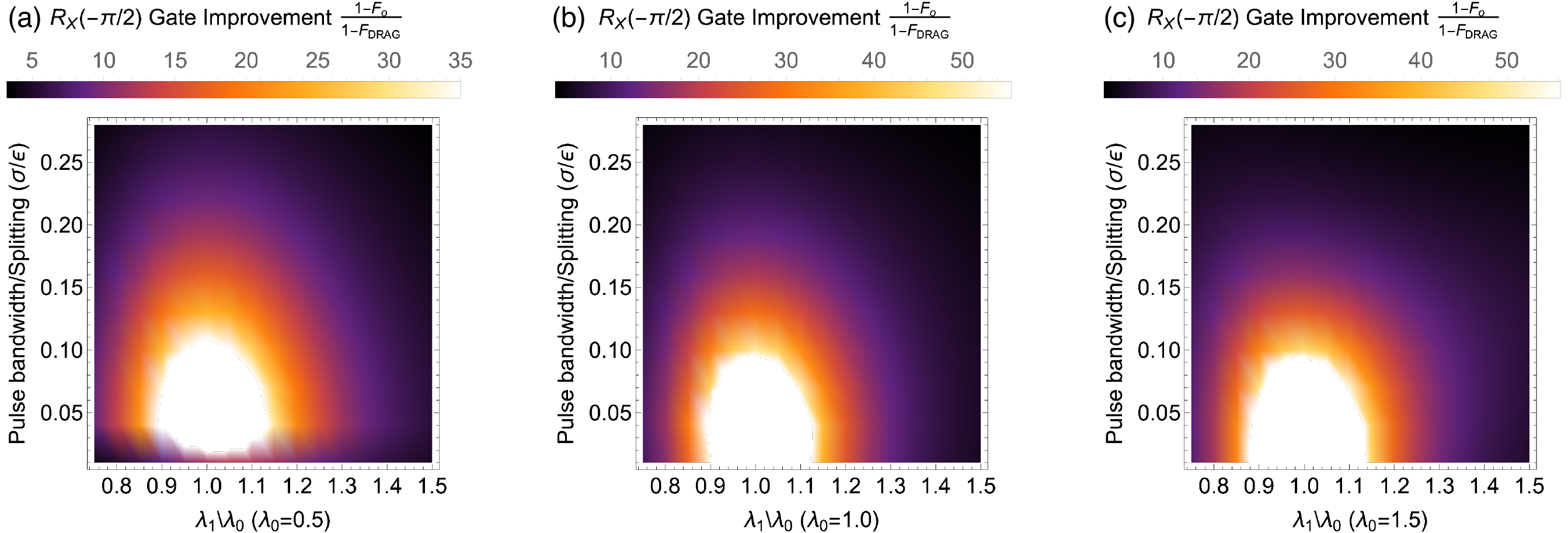}
\caption{(Color online)~Gate improvements for three different case of coupling strengths to the unwanted level, as a function of the bandwidth over splitting, with respect to the ratio of the unwanted couplings. The local minimas are due to the fact that at these certain values, the unitarity condition is satisfied therefore the DRAG formalism becomes more effective.}
\label{fig-DRAG-improvement}
\end{figure*}

Let us  discuss the validity of the approximation above. As discussed, the unwanted subspace in Eq.~\eqref{eq-first-order-drag} is neglected, since these elements constitute a higher-order error. This arises from the fact that in Eq.~\eqref{Eq7}, the $n$-th order DRAG Hamiltonian includes contributions from $S^{(n+1)}(t)$.  That is, the zeroth-order constraints ensure that $H_{\text{DRAG}}^{(0)}\equiv H_{\text{target}}^{\text{CPT}}$. The first-order target constraints give us the solutions for the pulse corrections (note that $H_{\text{DRAG}}^{(1)}$ includes contributions from $S^{(2)}(t)$, which for the purpose of this analysis we set arbitrarily as $S^{(2)}(t)=\textbf{0}$). We note that the first-order target constraints do not mix with higher elements of $S^{(2)}(t)$, which is not the case for the first-order decoupling constraints. Terminating at $n=1$ target constraints means that we leave second-order errors in the unwanted subspace of $H_{\text{DRAG}}^{(1)}$. If the goal is to find second-order pulse corrections, then one has to impose the second-order decoupling constraints, which will constrain certain elements of $S^{(2)}$ and disconnect the unwanted subspace of $H_{\text{DRAG}}^{(1)}$ from the qubit subspace. Then, applying the second-order target constraints on $H_{\text{DRAG}}^{(2)}$ gives rise to the second-order pulse corrections, by leaving 3rd order unwanted subspace errors in $H_{\text{DRAG}}^{(2)}$. This procedure can be repeated to obtain any subsequent order of pulse corrections as has been shown in Ref~\cite{Gambetta2011} for a transmon system. In the transmon case, the off-diagonal entries in the qubit subspace can be made to be zero in $H_{\text{DRAG}}^{(1)}$. In the $\Lambda$-system case, due to the different selection rules, we find that there is no transformation that can make the off-diagonal qubit subspace elements to be zero and simultaneously respect the boundary conditions of the transformation. Hence, attempting to go to higher-order pulse corrections will leave non-zero off-diagonal entries in the qubit subspace and different diagonal phase evolutions, which are difficult to incorporate analytically into the gate design. Consequently, we terminate the search of pulse corrections at the first order, and we show later on that such pulses can still lead to enhanced gate performance, without demanding higher-order modifications. The gate improvement is possible by resolving only the different phase evolution of the dark and bright-target subspaces, which makes our approximation for deeming the off-diagonal qubit subspace entries to be higher-order errors valid.    

We now highlight the procedure for finding the required detuning modification, on top of the pulse modulation, to correct for the phase error caused by the implementation of DRAG. We showcase this for the case of $\lambda_0=\lambda_1=1$ because, as we discussed above, it only leads to an induced phase between the bright and dark states up to the first order. To this end, we denote the ideal evolution operator that corresponds to $H_{\text{DRAG}}^{(0)}$ as $U_0$; this is the analytically solvable time evolution operator (which is also the ideal gate in the CPT frame). Our total Hamiltonian,  $H_{\text{DRAG}}=H_{\text{DRAG}}^{(0)}+H_{\text{DRAG}}^{(1)}$, evolves with a time evolution given by $U(t)=U_0(t)U(t)'$, and satisfies the equation:
\begin{equation}
     i\dot{U}_0U' +iU_0\dot{U}'=(H_{\text{DRAG}}^{(0)}+H_{\text{DRAG}}^{(1)})U_0U',
\end{equation}
which reduces to:
\begin{equation}
    i\dot{U}'=(U_0^\dagger H_{\text{DRAG}}^{(1)}U_0)U',
\end{equation}
where $U'$ is the evolution operator of $H_{\text{DRAG}}^{(1)}$ in the interaction picture of $H_{\text{DRAG}}^{(0)}$. Given the fact that the first-order error in $H_{\text{DRAG}}^{(1)}$ are the diagonal entries, we focus only on the bright-target subspace. In this subspace, $H_{\text{DRAG}}^{(1)}\propto f(t)\textbf{1}$, where $f(t)$ is the function that defines the relative phase shift between the dark and bright states. Solving the Schr$\ddot{\text{o}}$dinger equation we find that the induced phase is given by $\theta=-(\lambda_0^2+\lambda_1^2-6\lambda_0\lambda_1)\sigma/(4\varepsilon)$. Hence, in order to remove the $\theta$-shift from the target evolution we modify the detuning: 
\begin{equation} \label{eq-det-mod}
  \delta \to \delta' =\sigma/\tan(\frac{\phi+\theta}{2}),
\end{equation}
where $\phi$ is the  target rotation angle. The detuning modification~\eqref{eq-det-mod} together with Eqs. (\ref{eq-DRAG-sols-o}) and (\ref{eq-DRAG-sols-c}) complete our full set of solutions for the pulses. The final pulse shapes will depend on the specific details of the system such as the splitting of the two auxiliary states.

We demonstrate the performance of the DRAG solutions in Fig.~\ref{fig-comp-x}. The figure shows the fidelity of rotations about the $x$ axis, i.e. $R_{X}(-\pi)$ and $R_{X}(-\pi/2)$ gates, in terms of the dimensionless parameter given by the pulse bandwidth over the splitting between the target and unwanted levels. We consider the case in which both transitions are of equal strength larger than that with respect to the target level ($\lambda_0=\lambda_1=1.2$) and the case of unequally-strong couplings with one larger and one smaller than the coupling to the target state ($\lambda_0=0.8,~\lambda_1=1.2$). The top panels of Fig.~\ref{fig-comp-x} show the gate error of implemented operations and the bottom panels shows the deviation from unitarity of the operators, defined as $|UU^\dagger-\textbf{1}|$. It is clear from these plots that our DRAG correction provides substantial improvement compared to the uncorrected gate, in some cases by several orders of magnitude. It can also be seen that the operators do not remain unitary for the whole range of bandwidths over splittings, which indicates that the populations are leaving the qubit subspace. This signals that the sources of errors in this case are leakage, rather than phase errors, unlike the case of dependent couplings in Section~\ref{sec-exact-sol}. As such, our devised DRAG solution performs better at ranges of bandwidths over splittings where the gate implementation remains unitary, but it always provides an improvement throughout the full bandwidth range.

In Fig.~\ref{fig-DRAG-improvement}, we show the improvement in the fidelity of the $R_{X}(-\pi/2)$ gate as a function of both the bandwidth over splitting and the ratio of the unwanted couplings for three different values of these couplings with respect to the coupling to the target state (which is set to unity in all cases). In all cases, the best DRAG improvements occur at points where the values of the couplings are similar to each other, i.e. $\lambda_0 = \lambda_1$. The occurrence of minima near these values are related to the corresponding deviation from unitarity values: at these specific parameter values of the system, the transitionless condition is being satisfied and therefore the DRAG decoupling from the unwanted state becomes more efficient. Furthermore, notice that because our pulse corrections are functions of the couplings, improvements are better for the cases where the unwanted couplings are stronger compared to the coupling to the target state (i.e., $\lambda_{0(1)} > 1$). That is, for stronger couplings the corrections to the pulse shape will be more prominent and therefore DRAG becomes more effective. Consequently, in this formalism, we always pick the auxiliary state that has weaker dipole moments to be the target state.

\begin{figure*}[!htbp]
\includegraphics[scale=.77]{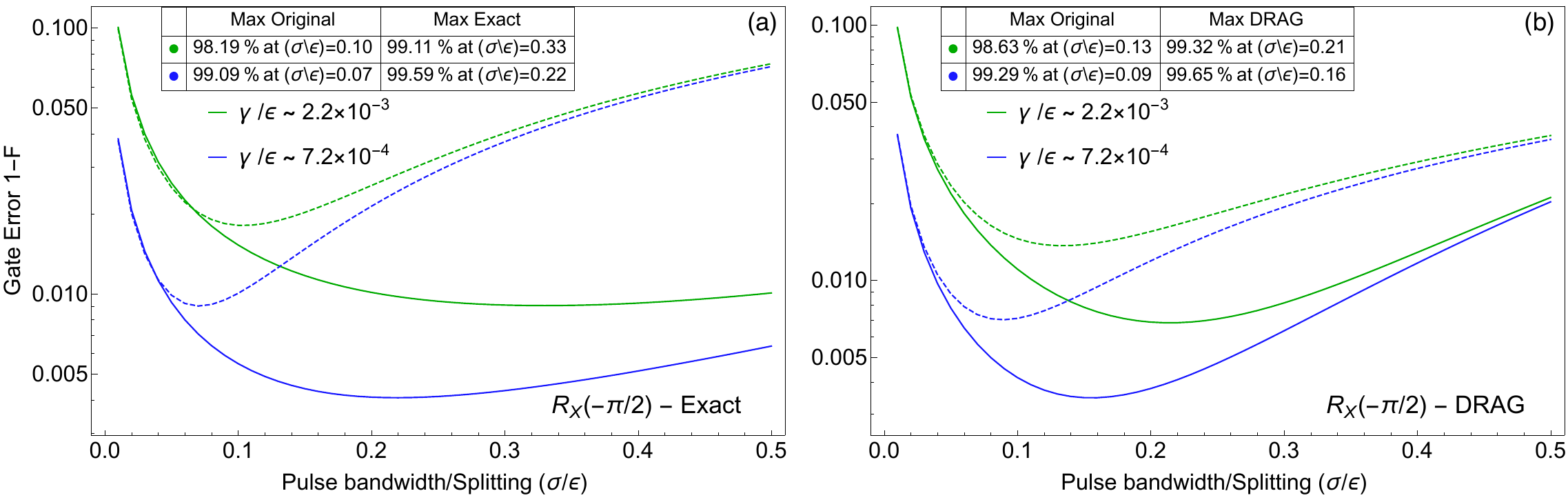}
\caption{(Color online)~The effects of spontaneous emission on $R_X(-\pi/2)$ using (a) exact with $\tan(\eta)=1.2$ and (b) DRAG solutions with $\lambda_0=1.2, \lambda_1=0.8$, for two different values of spontaneous emission in terms of the dimensionless ratio of the decay rate and the splittings $\gamma/\varepsilon$: $2.2\times 10^{-3}$ and $7.2\times 10^{-4}$. In each case the dashed lines are original pulses and the solid lines are the improved pulses. The inset values indicate the maximum fidelities and the corresponding bandwidths. With no corrective measures, best fidelities occur at narrow bandwidths. Our modified pulses lead to better fidelities at higher pulse bandwidths.}
\label{fig-SpE}
\end{figure*}

\section{Effects of additional errors} \label{sec-add-errors}

In this section, we consider the effects of errors other than the off-resonant couplings to the unwanted level to examine how the performance of our solutions is affected. In particular, we focus on two important sources of error: spontaneous emission from the excited levels and cross-talk between the transitions of the $\Lambda$-system.


\subsection{Spontaneous emission}\label{sec-se}

A realistic optically active system is coupled to the environment and therefore relaxes through the spontaneous emission of photons. This relaxation process reduces the fidelity of our intended quantum gates, more severely impacting longer gates. We describe the dynamics with a standard open quantum system approach to model the effects of spontaneous emission. We use the Liouville–von Neumann equation with Lindblad relaxation terms: $\dot{\rho}=-i\left[H, \rho\right]+\mathcal{L}[\rho]$, where $
\mathcal{L}[\rho]=\sum_{i j}\left(L_{i j} \rho L_{i j}^{\dagger}-\frac{1}{2}\left[L_{i j}^{\dagger} L_{i j} \rho+\rho L_{i j}^{\dagger} L_{i j}\right]\right)$. The Lindblad operators account for the populations that leave the target and unwanted levels upon emission of a photon. We pick four Lindblad operators that correspond to emission from each excited state to each ground state: $L_{ij} =\sqrt{\gamma}~| i\rangle\langle  j|$, where $i\in \{0,1 \}$,  $j\in \{t,u\}$, and $\gamma$ is the emission rate, which we take to be the same for all transitions. 

We have considered the effects of spontaneous emission for both cases of dependent and independent couplings. Fig.~\ref{fig-SpE} shows the effects of spontaneous emission on the corrected exact and DRAG solutions for different dimensionless values of $\gamma/\varepsilon$, for $R_X(-\pi/2)$. As evident from the figure, we see that two opposing effects are acting on the system. On one hand, the detrimental effect of the unwanted level favors short-bandwidth pulses. On the other hand, the spontaneous emission favors large-bandwidth pulses (i.e.\@ short pulses in time). This results in a minimum gate error for an intermediate value of the pulse bandwidth. (In the DRAG case, notice that this is independent of deviation from unitarity condition discussed in the previous section.) The insets in Fig.~\ref{fig-SpE} show the maximum obtainable value of fidelities with their corresponding values of bandwidth over splitting. For both values of $\gamma/\varepsilon$, while the corrective solutions lead to a better maximum fidelity, they also occur at larger values of bandwidth over splitting. Therefore our formalism provides us with better gate performances, at more reasonable values of pulse bandwidths, suitable for systems with fast spontaneous emission rates.

\begin{figure*}[!htbp]
\includegraphics[scale=.07]{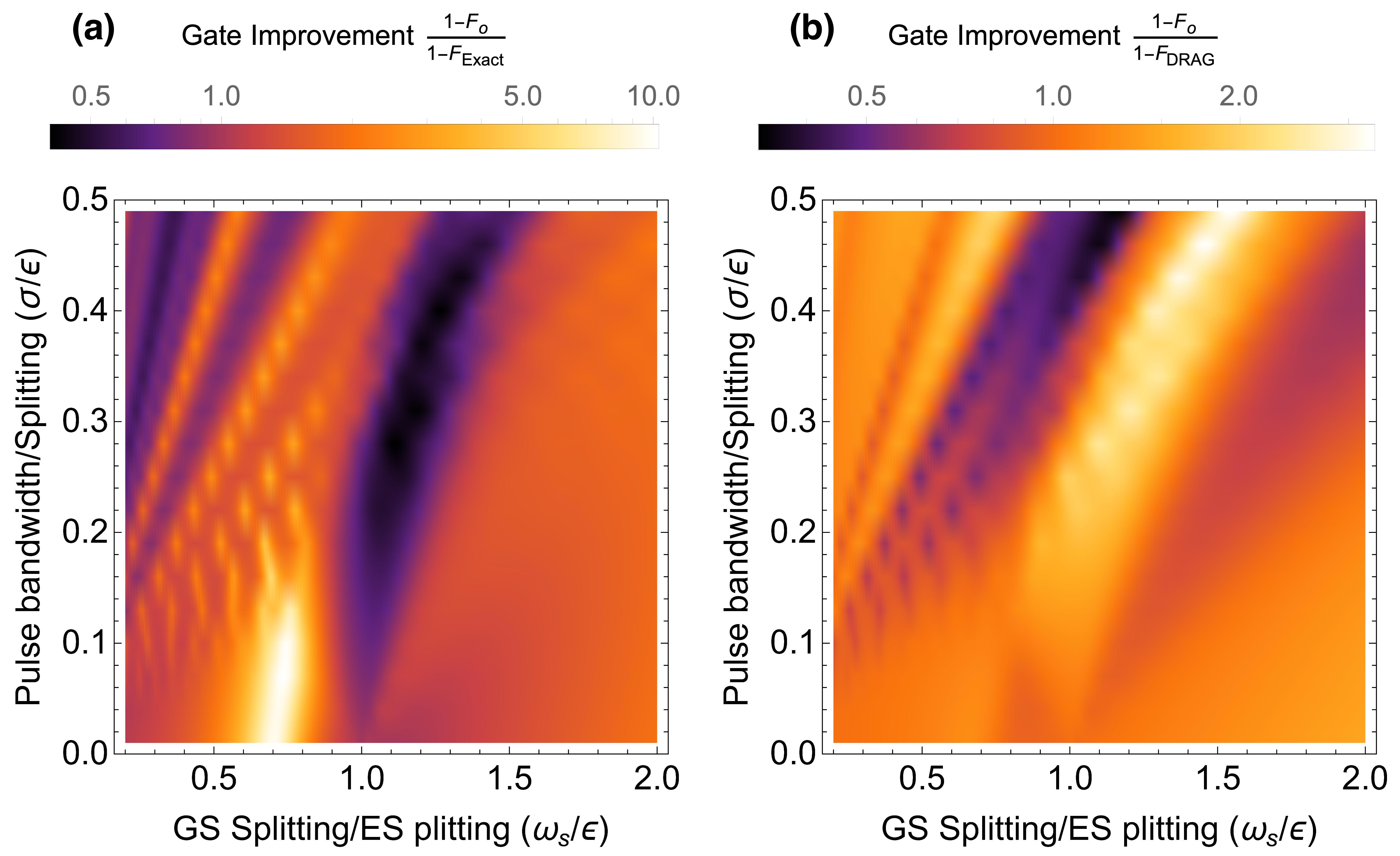}
\caption{(Color online)~Gate improvement of $R_X(-\pi/2)$ for (a) the exact method and (b) using the DRAG method, including the cross-talks between the two transitions of the $\Lambda$-system. The vertical axis is the dimensionless parameter pulse bandwidth over the excited states splitting ($\sigma/\varepsilon$), and the horizontal axis is the dimensionless parameter ground states splitting over the excited states splitting ($\omega_s/\varepsilon$). The other parameter values are $\eta=\pi/4$ for the exact method, and $\lambda_0=1,~\lambda_1=1$ ($\kappa_{01}=1=1/\kappa_{10}$)}
\label{fig-ct}
\end{figure*}

\begin{figure*}[!htbp]
\includegraphics[scale=.08]{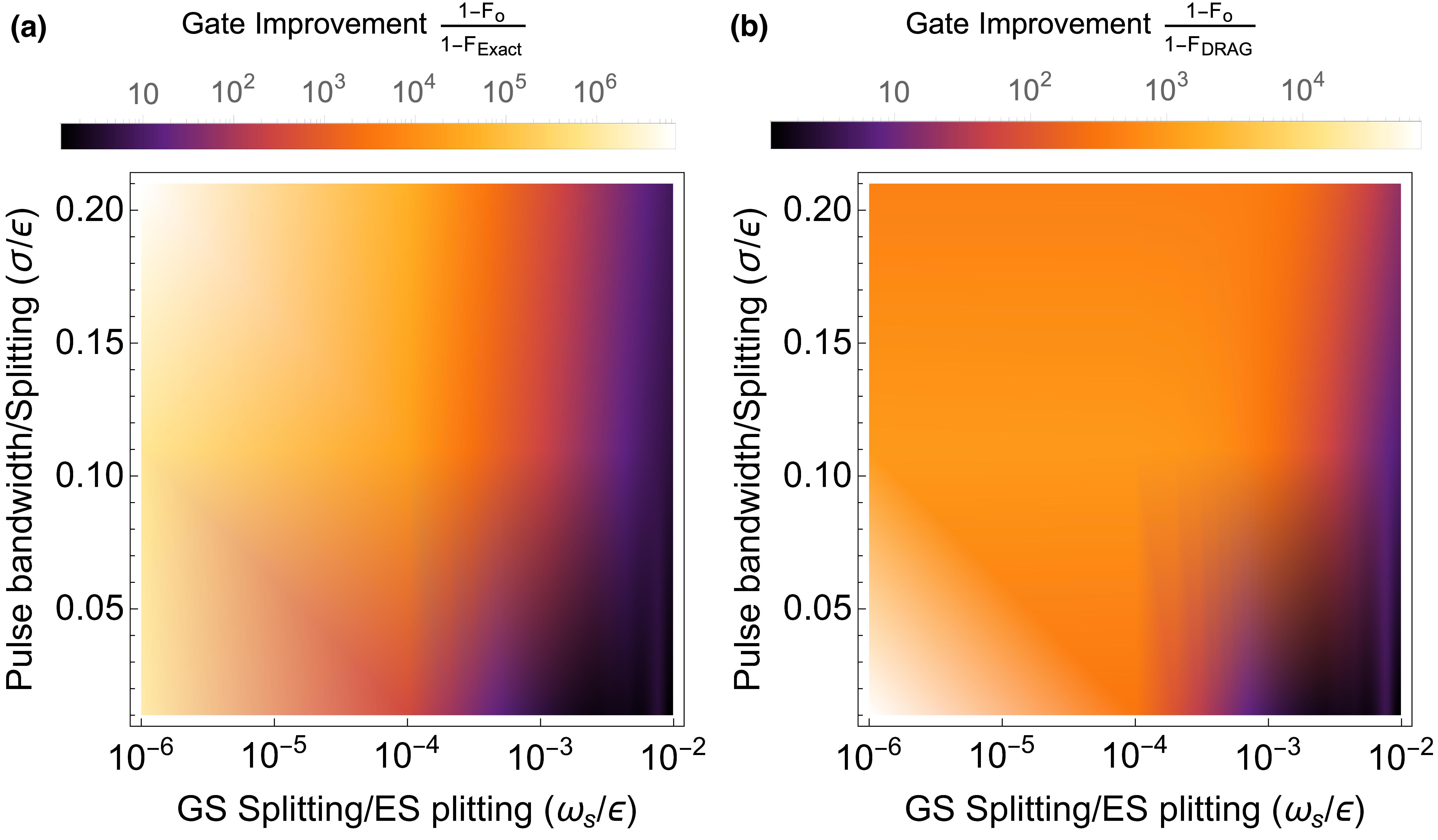}
\caption{(Color online)~$R_X(-\pi/2)$ gate improvement with cross-talks in the regime that $\varepsilon\gg \omega_s$ for (a) the exact method and (b) the DRAG method. In addition to the corrective measures, the Rabi frequencies are also multiplied by $1/2$ due to the cross-talk modifications to the system Hamiltonian (see the text). The other parameter values are $\eta=\pi/4$ for the exact method, and $\lambda_0=1,~\lambda_1=1$ ($\kappa_{01}=1=1/\kappa_{10}$).}
\label{fig-ct-improved}
\end{figure*}


\subsection{Cross-talk} \label{sec-crosstalk}

So far in our discussion we have assumed that either the polarization selection rules or the energy separation of the ground states allow to distinguish the two transitions of the $\Lambda$-system. Here we include the error from cross-talk of the two transitions. We model our interaction picture cross-talk Hamiltonian (for an $X$-rotation) in the frame rotating with the frequency of the pulses, as follows:
\begin{eqnarray} \label{eq-H-CT}
H_{\text{CT}}&=&(\delta/2) (\Pi_0 +\Pi_1-\Pi_t-\Pi_u)+\varepsilon \Pi_u \nonumber  \\
&&+~ \big\{ (1/2)(\Omega_o -i\Omega_c)(1+e^{-i\omega_s t}\kappa_{10})|0\rangle\langle t| 
\nonumber\\
&&+~(1/2)(\Omega_o -i\Omega_c)(\lambda_0+e^{-i\omega_s t}\lambda_0\kappa_{10})|0\rangle\langle u| 
\nonumber\\
&&+~(1/2)(\Omega_o -i\Omega_c)(1+e^{-i\omega_s t}\kappa_{01})|1\rangle\langle t| 
\nonumber\\
&&+~(1/2)(\Omega_o -i\Omega_c)(\lambda_1+e^{-i\omega_s t}\lambda_1\kappa_{01})|1\rangle\langle u| \nonumber\\
&&+~ \text{h.c.}\big\}.
\end{eqnarray} Here $\omega_s$ quantifies the ground state (GS) splitting (i.e., the splitting between the qubit states, $\ket 0$ and $\ket 1$, in the lab frame). Following the assumption on the transition dipoles discussed in Section~\ref{sec-system}, the cross-talk couplings may also be
implicitly dependent or independent of each other. When the $E_0$ ($E_1$) field drives the $\ket 1 \to \ket t$ ($\ket 0 \to \ket t$) transition, it leads to cross-talk coupling $\kappa_{01}=E_0/E_1$ ($\kappa_{10}=E_1/E_0$). Consequently, once the $E_0$ field drives the $\ket 1 \to \ket u$ transition, the resulting cross-talk coupling will be $\lambda_1\kappa_{01}$, and similarly, the effect of $E_1$ driving the $\ket 0\to \ket u$ transition will be the coupling $\lambda_0\kappa_{10}$.

Fig.~\ref{fig-ct} shows the effect of cross-talk on the gate improvement of $R_X(-\pi/2)$ for both the exact and the DRAG method. The values are shown for a range of dimensionless parameters of pulse bandwidths over excited state (ES) splittings ($\sigma/\varepsilon$), and GS splitting over ES splitting ($\omega_s/\varepsilon$). In (a), for the exact method, we have set $\eta=\pi/4$ and in (b), for the DRAG method, we have set $\lambda_0=1,~\lambda_1=1$. Furthermore, the ratio of the two electric fields is chosen to be unity $E_1/E_0=1$ (i.e., $\kappa_{01}=1=1/\kappa_{10}$). As the figure indicates, the behavior of the improvement heavily relies on the specific parameters of the system. For example, the cross-talk Hamiltonian corresponding to Fig.~\ref{fig-ct}(a) (i.e., $\lambda_0=-1,~\lambda_1=1$) in the CPT frame after RWA will be,

\begin{equation}
\begin{split}
    \tilde{H}_{\text{CT},\text{CPT}}&=(\delta/2)(\Pi_B+\Pi_D-\Pi_t-\Pi_u)+\varepsilon\Pi_u \\&~+ \frac{1}{\sqrt{2}}\Big\{ i(1+\cos(\omega_st)) \sum_{j=o,c}\Omega_j\sigma^j_{B,t}  \\&~+
    (-i)\sin(\omega_st) \sum_{j=o,c}\Omega_j\sigma^j_{D,t} 
    \\&~+(-i)\sin(\omega_st)\sum_{j=o,c}\Omega_j\sigma^j_{B,u} \\&~+i(1+\cos(\omega_st))\sum_{j=o,c}\Omega_j\sigma^j_{D,u} \Big \}.
    \end{split}
\end{equation}

This indicates that, firstly, the pulse profiles will be modulated with some time-dependent, periodic functions of $\omega_s$, therefore the pulse behavior will be non-trivial. Additionally we notice that for $\varepsilon\gg \omega_s$, the system reduces to that of two two-level systems. This means that in this regime our exact solution is applicable. Similarly, looking at the corresponding Hamiltonian to Fig.~\ref{fig-ct}(b) (i.e., $\lambda_0=1,~\lambda_1=1$) we find that in the same $\varepsilon\gg \omega_s$ limit, it reduces to that of the V-system for which we developed the DRAG formalism. Both of these cases require a slight correction of Rabi frequencies by a factor of 1/2 to account for additional factor of 2 arising from the terms $1+\cos(\omega_st)$. We show the performance of these solutions in the presence of cross-talk in Fig.~\ref{fig-ct-improved}. As seen in both figures, improvements of several orders of magnitude are achievable through our approaches given that the condition $\varepsilon\gg \omega_s$ is satisfied.  
We also note that in the case of dependent couplings, because the errors are in the form of phase errors, in principle it is possible to track the phase change and correct it through detuning modification. Moreover, for a generic DRAG-based approach to avoid the cross-talks in $\Lambda$-systems we refer to the work in Ref.~\cite{takou2021}.

\begin{figure*}[!ht]
\includegraphics[scale=.7]{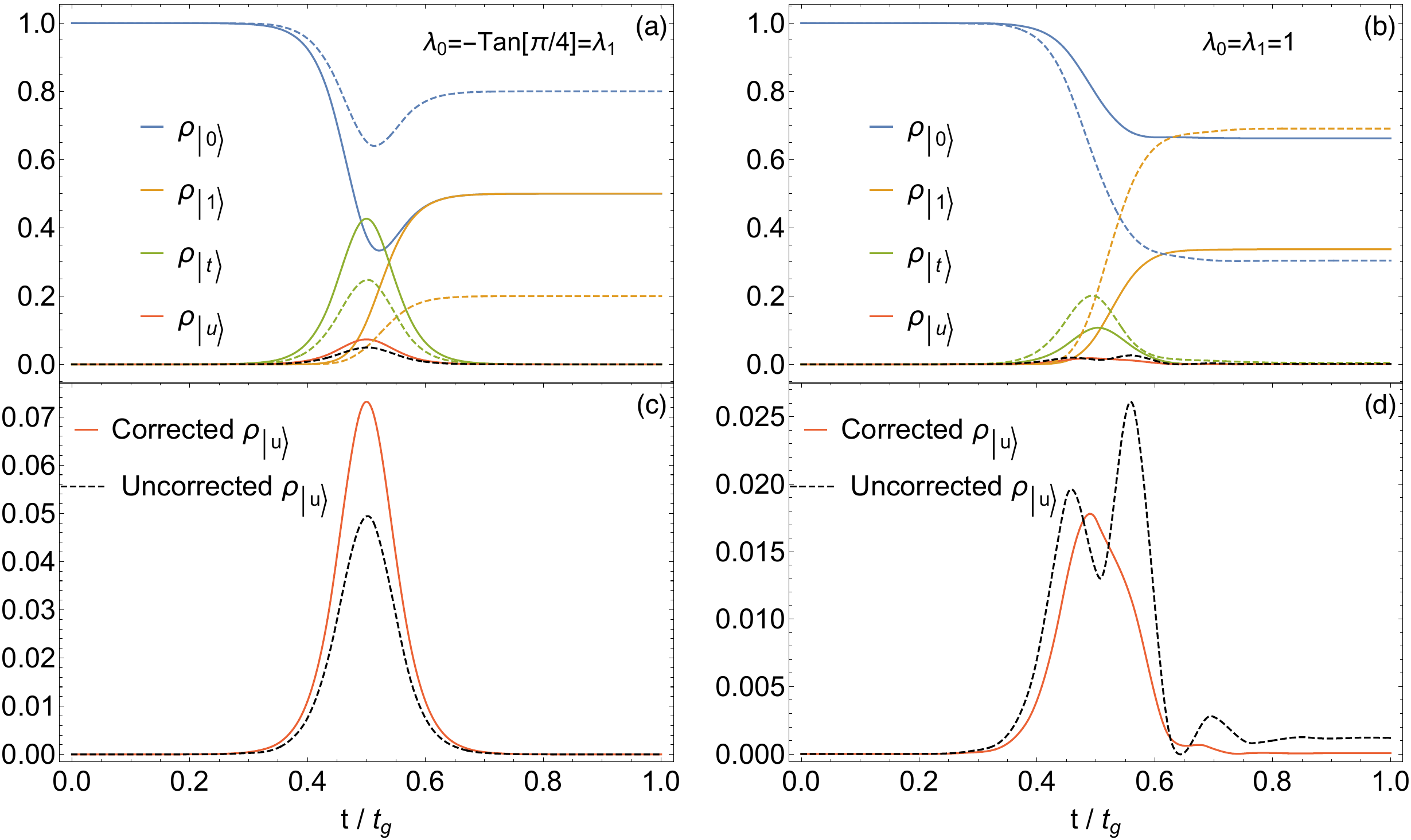}
\caption{(Color online)~The dynamics of population during $R_X(-\pi/2)$ gate for all four levels of a $\Lambda$-system with an unwanted level. In (a) we have used the exact method (for dependent couplings), and in (b) we have used the DRAG method (for independent couplings of same sign). The bandwidth over splitting ratio is taken to be $\sigma/\varepsilon=0.5$. The dashed curves indicate the populations of a non-modified system and the solid lines indicate populations of a system subject to our modified solutions. Bottom panels (c) and (d), are specifically focused on the population of the unwanted level for (a) and (b), respectively. The populations of target and unwanted levels subject to a transitionless pulse go back to the qubit states at the end of the gate time in (a) and (c). However, in the case of dependent couplings in (b) and (d), we deal with leakage to unwanted level. Both of our designed formalisms correct the population of the qubit levels with respect to desired gate, and additionally, our DRAG method reduces the leakage in (d).}
\label{fig-pop}
\end{figure*}


\section{Discussion and Conclusion} \label{sec-conclusion-drag}

In this work we have developed a framework for high-fidelity control of $\Lambda$-systems with unwanted transitions for two cases of dependent and independent couplings to the unwanted level. The case of the dependent couplings is motivated by the fact that off-resonant couplings are likely to arise due to the physical origin of the excited states as orthogonal superpositions of two basis states. For example, in quantum dot molecules where the qubit states are the spinor projection of a single hole, both the $\ket{t}$ and $ \ket{u} $ states are created by the bonding and anti-bonding superpositions of two different charge and spin configurations, with the coupling mediated by tunneling. In such an example, the degree of mixing can be controlled by applied electric fields~\cite{Economou2012}. 
We showed that in this case the $\Lambda$-system reduces to a combination of a two two-level-systems in the limit of equal basis weights ($\eta=\pi/4$). We further discussed that the errors are in the form of phase errors; as such we devised an exact solution that compensates for these errors in the form of detuning modification and allows for error-free gate implementations. On the other hand, in the absence of basis states, the system turn into a V-system in the limiting case of equal couplings with the same sign. In this system the source of errors has the form of leakage rather than phase errors. In Section~\ref{sec-DRAG} we developed a novel version of the DRAG approach to battle the leakage in this case.

To see the different form of errors (i.e., phase error versus leakage), we briefly comment on the evolution of population of each level during the gate time. In Fig.~\ref{fig-pop} we have shown the populations of the system for a $R_X(-\pi/2)$ rotation, for the cases of both dependent and independent couplings. The transitionless nature of the pulses driving two independent two-level-systems is evident in Fig.~\ref{fig-pop}(c): the final population of excited states are negligible for both uncorrected and corrected solutions. This indicates that the core of the errors in our system are the phase errors due to the off-resonant couplings to unwanted states, rather than leakage. Our formalism corrects these errors by populating the excited states appropriately and redistributing the populations on the qubit states such that the desired gates are implemented. 
On the other hand, as we discussed in Section~\ref{sec-DRAG}, in the independent couplings case the evolution of the system is nonunitary. As such, by the end of the gate time some of the population leaves the qubit subspace, being stored in the unwanted level (Fig.~\ref{fig-pop}(d)). Our DRAG approach battles this leakage through modulation of the pulses and restores the full population to the qubit subspace. However, this comes at the price of inducing some phase errors which are compensated through the additional detuning modification.

In summary, the two methods we present above are capable of implementing high fidelity arbitrary gates in either case of a $\Lambda$-system with unwanted transitions. The exact method has the advantage that it only requires a simple modification of the detuning and therefore is less expensive from an experimental point of view. The DRAG pulse modification, on the other hand, is in the form of a corrective modification to the original pulse shape that drives each transition of the system.

It should be noted that the DRAG pulse modulation solution is not unique; we have made specific choices on the generative parameters of the DRAG frame in deriving our DRAG solutions (see Appendix~\ref{app-derivations} for details of the derivations). These generative elements are the free parameters of the system and can be set to arbitrary values as long as they satisfy the DRAG transformation condition $S(t_g)=S(0)=0$. Therefore, alternative DRAG solutions based on different generative elements can also be devised. We have demonstrated such pulse shapes in Fig.~\ref{fig-pulse-shape}. The corrective pulse is inversely proportional to the splitting $\varepsilon$ and the coupling to the unwanted level for that transition (Eq.~\eqref{eq-DRAG-sols-c}). For smaller splittings, the correction pulse becomes more comparable to the original pulse. We note that even though the mathematical derivation of the pulse design is phrased in terms of an original (uncorrected) pulse and a correction, experimentally the full corrected pulse would be programmed and generated directly. 

Finally, we discussed the implementation of $X$-, and $Y$-rotations. To achieve universal control of the qubit system, we also require implementations of $Z$-rotations. This can be done either through the same control scheme by driving only a single transition of the $\Lambda$-system with a sech pulse~\cite{Economou2006} or simply through `virtual' $Z$ gates \cite{McKay2017PRA}. 
The exact solution and the version of DRAG we have tailored to the case of a $\Lambda$-system with a fourth unwanted level are general and can be applied to a variety of optically active qubits, such as color centers (e.g., the NV center in diamond), trapped ions, rare-earth ions, self-assembled quantum dots, and quantum dot molecules.




\section*{ACKNOWLEDGMENTS}

This work was supported by the NSF. M.D. and S.E.E. acknowledge Grant No. 1839056. S.E.E. also acknowledges support from Grant No. 1838976.


\appendix


\section{Coherent population trapping} \label{app-cpt}

In this appendix, we present the mathematical details of the CPT framework. Two-level systems subject to a sech pulse with Rabi frequency $\Omega$ and bandwidth $\sigma$, can be solved analytically~\cite{Rosen} and the solutions are in the form of hypergeometric functions. For the case of $\Omega/\sigma \in \mathbb{N},$ these pulses are transitionless~\cite{Economou2006}, i.e., after the passage of the pulse the population will always return to the ground state with the ground state acquiring a non-trivial phase $\phi$ (Eq.~\eqref{eq-phase}) through the process. For the specific case of $\Omega=\sigma$, the Hamiltonian of a generic two-level system driven by a sech pulse in the rotating frame is given by

\begin{equation}
    H = \begin{pmatrix} 0 & \Omega(t) e^{-i\delta t}\\
    \Omega(t)e^{i\delta t} & 0
    \end{pmatrix},
\end{equation}
where $\Omega(t)= \sigma \sech(\sigma (t - t_g/2))$. For this system the unitary evolution operator at the end of the gate is $U=\text{diag}(e^{- i\phi},e^{i \phi})$. In the CPT scheme, the transitions of the system are excited using the drive field, $E_0 f_0(t) e ^{i \omega_0 t} +e^{i \alpha} E_1 f_1(t) e ^{i \omega_1 t} +\text{h.c.}$ For identical temporal envelopes and detunings, with Rabi frequencies $\Omega_0(t)$ and $\Omega_1(t)$, the transformation of the original qubit states to the bright and dark states is given by:

\begin{equation} \label{eq-CPT-transformation}
\begin{pmatrix}
\ket{D} \\ \ket{B}
\end{pmatrix} = \begin{pmatrix}
\cos \frac{\theta}{2} & -e^{i \alpha} \sin\frac{\theta}{2} \\ e^{-i\alpha}\sin\frac{\theta}{2}   & \cos\frac{\theta}{2} 
\end{pmatrix}\begin{pmatrix}
\ket{0} \\ \ket{1}
\end{pmatrix},
\end{equation}

where $\sin\frac{\theta}{2} = \Omega_0/\Omega_\text{eff}$, $\cos\frac{\theta}{2} = \Omega_1/\Omega_\text{eff}$, and $\Omega_\text{eff}^2 = \Omega_0^2 + \Omega_1^2$. In the CPT frame, the transition matrix elements between the target and the dark state vanishes and the bright and target state will have the matrix element defined by the effective Rabi frequency: $V_{t,B} =\Omega_\text{eff} f(t) e^{- i \delta t}$. For the case of both drives using a sech temporal envelope, i.e., $f_0(t)=f_1(t)=\sech(\sigma t)$, the transitionless pulse with $\Omega_\text{eff} = \sigma$ will induce the relative phase $\phi$ between the bright and dark states which translates to a rotation in the subspace of $\ket D$ and $\ket B$. Therefore, by varying the drive parameters we can set the unitary transformation of Eq.~(\ref{eq-CPT-transformation}) to transform our original qubit states to the desired states in the CPT frame, effectively enabling the rotation about an arbitrary axis of rotation $\hat n = (\sin \theta\cos\alpha,\sin\theta\sin\alpha,\cos\theta)$: $R_n(\phi) = e^{-i \phi \hat n . \vec \sigma}$.

The CPT framework can be applied to the case of $\Lambda$-system with unwanted level in a similar way. However, there will be additional transitions from the bright and dark state to the unwanted level. In the lab frame of a non-ideal system with an unwanted level, for the case of equal detunings $\delta$, the Hamiltonian in the interaction frame after the RWA can be written as
\begin{eqnarray} \label{eq-X-H-int}
H_\text{int}&=&  \sum_{j=0,1} e^{i (\delta-\varepsilon) t}  \lambda_j \Omega_{j}  |j\rangle\langle u| \nonumber \\
&&+  \sum_{j=0,1} e^{i \delta t} \Omega_{j}  |j\rangle\langle t|
+ \text{h.c.}  
\end{eqnarray}
The CPT transformation for an $X$-rotation amounts to having both Rabi frequencies to be equal: $\Omega_o(t)=\Omega_{0}(t)=\Omega_{1}(t)$ (we set $\theta=\pi/2$ and $\alpha=0$). Such CPT transformation turns this Hamiltonian into
\begin{eqnarray} \label{eq-X-H-CPT-no-drag}
H_\text{CPT}&=&e^{i \delta t} \sqrt{2}~\Omega_o |B\rangle\langle t| \nonumber \\
&&+~ \frac{e^{i (\delta-\varepsilon) t }}{\sqrt{2}} \big\{\Omega_o (\lambda_0 -\lambda_1) |D\rangle\langle u| \nonumber \\
&&+~\Omega_o (\lambda_0 +\lambda_1)  |B\rangle\langle u|\big\}+ \text{h.c.} 
\end{eqnarray}
We proceed by removing the oscillatory parts of the CPT Hamiltonian by going to a rotating frame. We do this using the frame transformation
\begin{equation}
    \text{diag}\big[e^{-i (\delta/2) t},e^{-i (\delta/2) t},e^{i (\delta/2) t},e^{i (\delta/2-\varepsilon) t} \big].
\end{equation}

Upon this transformation we arrive at the Hamiltonian given by Eq.~\eqref{eq-X-H} of the main text. Repeating the same series of transformations outlined above, but now this time with the additional controls $\Omega_{\ell,c}~(\ell=0,1)$ under the condition $\Omega_o\equiv{\Omega}_{0,o}={\Omega}_{1,o}$, and $\Omega_c\equiv{\Omega}_{0,c}={\Omega}_{1,c}$, leads to the Hamiltonian given in Eq.~\eqref{eq-X-H-DRAG}.


\section{Mathematical derivation of DRAG solutions} \label{app-derivations}

In this appendix, we lay out the details of our DRAG-based formalism that leads to the solutions presented in Eqs.~\eqref{eq-DRAG-sols-o} and \eqref{eq-DRAG-sols-c}. To that end, we employ the perturbative DRAG theory developed in Ref.~\cite{Gambetta2011} as the base of our formalism. As our first step, we need to expand the control fields of the DRAG frame Hamiltonian with respect to  the adiabatic parameter $x$:
\begin{equation} \label{eq-Heff}
H_\mathrm{DRAG}^{(n)}(t)=H_\mathrm{extra}^{(n)}(t)+\bar{H}^{(n)}(t) + i [S^{(n+1)}(t),\Pi_u],
\end{equation}

where $n\geq 0$ corresponds to the order of the transformed Hamiltonian, and $H_\mathrm{extra}$ is a nontrivial expression generated by the lower orders of the transformation, and,

\begin{equation}
   \bar{H}_{\text{CPT},\omega d}(t) = \frac{1}{x} \Pi_u + \sum_{n=0}^\infty x^n \bar{H}^{(n)}(t).
\end{equation}

Notice that this expansion essentially means that the constraints in Eqs.~(\ref{eq-X-control}) and (\ref{eq-X-decoupling}), and consequently the control fields $h_i(t)$ and $h_z(t)$ should be made perturbative with respect to the order of this expansion as well. Furthermore, the general form of the Hermitian operator $S(t)$ for a $d$-dimensional system can be written as

\begin{equation}
    S(t)=\sum_i s_{i,z}(t) \Pi_i + \sum_{i=o,c}\sum_{m<n} s_{i,m,n}(t)\sigma_{m,n}^i.
\end{equation}

The zeroth- and first-order expressions of $H_\mathrm{extra}$ are as follows \cite{Gambetta2011}:
\begin{widetext}

\begin{eqnarray} \label{eq-H-extra}
H_\mathrm{extra}^{(0)} &=& 0, \nonumber \\
H_\mathrm{extra}^{(1)} = i [S^{(1)}(t),H^{(0)}(t)] -&\frac{1}{2}& [S^{(1)}(t),[S^{(1)}(t),\Pi_u]] -\dot{S}^{(1)}(t)
\end{eqnarray}

\end{widetext}

Using Eqs.~(\ref{eq-Heff}) and \eqref{eq-H-extra}, we can solve for the constrains in Eqs.~(\ref{eq-X-control}) and (\ref{eq-X-decoupling}) to obtain the appropriate control elements in terms of different orders of the parameter $x$. The control constraints of Eq. \eqref{eq-X-control} turn into
\begin{eqnarray}\label{eq-constraints-X-Control}
\sqrt{2} \bar \Omega_o^{(n)} &=& h^{(n)}_o-\mathrm{Tr}[ H^{(n)}_\mathrm{extra}(t) \sigma^o_{B,t}], \nonumber \\
 \sqrt{2} \bar \Omega^{(n)}_{c}& =& h^{(n)}_c-\mathrm{Tr}[ H^{(n)}_\mathrm{extra}(t) \sigma^c_{B,t}],  \\
\bar\delta^{(n)}(t) &=&h_z^{(n)}(t)- \mathrm{Tr} [ {H}_\mathrm{extra}^{(n)}(t)\left(\Pi_B -  \Pi_t \right)]. \nonumber 
\end{eqnarray}
Note that we have picked the two original drives to have the same pulse envelope $\Omega_{0,o}(t)=\Omega_{1,o(t)}\equiv\Omega_{o}(t)$, and we have set $\Omega_{0,c}(t)=\Omega_{1,c}(t)\equiv\Omega_{c}(t)$. This will implement an $X$-rotation. A $Y$-rotation can be implemented in a similar manner, except that the two original drives will have a $\pi/2$ phase difference. The phase difference, however, will not show up in the CPT frame and thus, the rest of the analysis will be identical in both cases. The decoupling constraints of Eq.~\eqref{eq-X-decoupling} turn into 

\begin{eqnarray} \label{eq-decoupling-n-X}
   s_{c,D,u}^{(n+1)} &=&-\frac{1}{2} \mathrm{Tr} [ {H}_\mathrm{extra}^{(n)}(t) \sigma^o_{D,u} ]-
\frac{1}{2\sqrt{2}}(\lambda_0-\lambda_1)\bar \Omega_o^{(n)}, \nonumber \\
 s_{o,D,u}^{(n+1)} &=&+\frac{1}{2} \mathrm{Tr} [ {H}_\mathrm{extra}^{(n)}(t) \sigma^c_{D,u} ]+
\frac{1}{2\sqrt{2}}(\lambda_0-\lambda_1)\bar\Omega_{c}^{(n)}, \nonumber \\
  s_{c,B,u}^{(n+1)} &=&-\frac{1}{2} \mathrm{Tr} [ {H}_\mathrm{extra}^{(n)}(t) \sigma^o_{B,u} ]-
\frac{1}{2\sqrt{2}}(\lambda_0+\lambda_1)\bar \Omega_o^{(n)}, \nonumber \\
 s_{o,B,u}^{(n+1)} &=&+\frac{1}{2} \mathrm{Tr} [ {H}_\mathrm{extra}^{(n)}(t) \sigma^c_{B,u} ]+
\frac{1}{2\sqrt{2}}(\lambda_0+\lambda_1)\bar\Omega_{c}^{(n)}, \nonumber \\
s_{c,t,u}^{(n+1)} &=&-\frac{1}{2} \mathrm{Tr} [ {H}_\mathrm{extra}^{(n)}(t) \sigma^o_{t,u} ], \nonumber \\
 s_{o,t,u}^{(n+1)} &=&+\frac{1}{2} \mathrm{Tr} [ {H}_\mathrm{extra}^{(n)}(t) \sigma^c_{t,u} ]. 
\end{eqnarray}

From these constraints, we first find the zeroth-order solutions. According to the CPT framework, in order to implement $X$-rotations we set $h_{o}^{(0)}(t)=\sqrt{2}~ t_g\Omega(t)$ where $\Omega(t)=\sigma\sech(\sigma t)$, $h_{c}^{(0)}(t)=0$, and $h_z^{(0)}(t)=t_g\delta$. This implies that the target CPT Hamiltonian has the form:
\begin{equation}
   H_\mathrm{target}^{\mathrm{CPT}}=[\sigma\sech(\sigma t)\sigma_{B,t} +\text{h.c.}]
\end{equation}
Note that the higher orders of $h_{o}^{(n)}$ and $h_c^{(n)}$ are set to zero, such that we obtain the desired evolution as dictated by the target Hamiltonian. Since we made our choice for the control fields $h(t)$, by making use of $H_{\text{extra}}^{(0)}=0$ we can now solve for the pulse envelopes, from which we find:
\begin{equation}
    \Omega_{o} (t)= \sqrt{2}~ \Omega(t),\enskip \Omega_{c} = 0.
\end{equation}
This ensures that we have the right form in the bright-target subspace of $H_{\text{DRAG}}^{(0)}$ to match the corresponding bright-target elements of the target Hamiltonian. (Notice that the additional factor of $\sqrt{2}$ is present due to the CPT frame transformation.) Next, we proceed with the zeroth-order decoupling constraints. These constraints together with the zeroth-order target constraints will fix certain elements of $S^{(1)}(t)$, such that we  essentially satisfy $H_{\text{DRAG}}^{(0)}\equiv H_\mathrm{target}^{\mathrm{CPT}}$. Given the fact that $H_{\text{extra}}^{(0)}=0$, we find based on Eqs.~(\ref{eq-decoupling-n-X}) that the non-zero elements of $S^{(1)}(t)$ are $s_{c,D,u}^{(1)} = -\frac{1}{2\sqrt{2}}(\lambda_0-\lambda_1) t_g \Omega(t), \quad \text{and} \quad s_{c,B,u}^{(1)} = - \frac{1}{2\sqrt{2}}(\lambda_0+\lambda_1) t_g \Omega(t)$. These two constraints ensure no transitions between dark-unwanted and  and bright-unwanted states, respectively. The first order corrective controls are found by satisfying the first-order target constraints. In this first order, as we already mentioned we set all the first-order control fields to zero. Making use of Eqs.~\eqref{eq-constraints-X-Control}, we find the following equations:
\begin{eqnarray}\label{eq-corrections-list-x}
\sqrt{2} \bar\Omega_{o}^{(1)} &=& 2 \dot{s}_{o,B,t}^{(1)}+2 \delta  s_{c,B,t}^{(1)},  \\
\sqrt{2}\bar\Omega_{c}^{(1)} &=& 2 \left(\dot{s}_{c,B,t}^{(1)}+\sqrt{2}\Omega  t_g (s_{z,D}^{(1)}- s_{z,B}^{(1)})-2\delta s_{o,B,t}^{(1)} \right),\nonumber \\
s_{c,B,t}^{(1)} &=& \frac{ 1}{16\sqrt{2} } [\left(\lambda _0+\lambda _1\right) ^2\Omega  t_g +8(\dot{s}_{z,D}^{(1)}-\dot{s}_{z,B}^{(1)})(\Omega  t_g)^{-1}].\nonumber 
\end{eqnarray}
To seek the simplest solution which satisfies the $S^{(1)}(0)=S^{(1)}(t_g)=0$ (such that the implemented gate is the same in the DRAG and CPT frame), we pick the free parameters $s_{z,i}^{(1)}$'s and $s_{o,B,t}^{(1)}$ all equal to zero. Substituting $s_{c,B,t}^{(1)}$ into the first two equations, we find the pulse corrections given in Eqs.~\eqref{eq-DRAG-sols-o} and~\eqref{eq-DRAG-sols-c} (notice that we require an additional factor of $\sqrt{2}$ in the non-CPT frame to accommodate for the CPT transformation). Substituting these solutions and the choices we made above for the generative elements of $S^{(1)}(t)$ in the expansion Eq.~\eqref{eq-Heff} will lead to the first-order form of the DRAG Hamiltonian $H_{\text{DRAG}}^{(1)}(t)$ given in Eq.~\eqref{eq-first-order-drag}. Note that since we terminate the expansion in the first order we set $S^{(2)}(t)=0$.


\bibliography{biblo}

\end{document}